\documentclass{elsarticle}

\usepackage{graphicx}
\usepackage{dcolumn}
\usepackage{bm}
\usepackage{amsmath,amssymb}
\usepackage{xspace}
\usepackage{xfrac}
\usepackage{color,soul}
\usepackage{graphicx}
\usepackage[colorlinks=true,urlcolor=blue,citecolor=blue,linkcolor=blue,bookmarks=false,pdfstartview={FitH}]{hyperref}
\usepackage{braket}
\usepackage{hyperref}

\newcommand{\br}{{\bf r}}
\newcommand{\bR}{{\bf R}}
\newcommand{\bk}{{\bf k}}
\newcommand{\bK}{{\bf K}}

\begin{document}

\title{Rare regions and avoided quantum criticality   in  disordered Weyl semimetals and  superconductors}

\author[add,add1,add2]{J. H. Pixley}
\ead{jed.pixley@physics.rutgers.edu}
\author[add]{Justin H. Wilson}
\ead{jhw81@physics.rutgers.edu}
\address[add]{Department of Physics and Astronomy, Center for Materials Theory, Rutgers University, Piscataway, NJ 08854, USA}

\address[add1]{Center for Computational Quantum Physics, Flatiron Institute, 162 5th Avenue, New York, NY 10010. }

\address[add2]{Physics Department, Princeton University, Princeton, New Jersey 08544, USA}

\begin{abstract}
Disorder in Weyl semimetals and superconductors is surprisingly subtle, attracting attention and competing theories in recent years.
In this brief review, we discuss the current theoretical understanding  of the effects of short-ranged, quenched disorder on the low energy-properties of three-dimensional, topological Weyl semimetals and superconductors. 
We focus on the role of non-perturbative rare region effects on destabilizing the semimetal phase and rounding the expected semimetal-to-diffusive metal transition into a cross over. 
Furthermore, the consequences of disorder on the resulting nature of excitations, transport, and topology are reviewed. 
New results on a bipartite random hopping model are presented that confirm previous results in a $p+ip$ Weyl superconductor, demonstrating that particle-hole symmetry is  insufficient to help stabilize the Weyl semimetal phase in the presence of disorder. 
The nature of the avoided transition in a model for a single Weyl cone in the continuum is discussed. 
We close with a discussion of open questions and future directions.
\end{abstract}

\maketitle

\tableofcontents

\section{Introduction}

The massive Dirac equation, first shown to explain the relativistic electron by Dirac~\cite{Dirac-1928}, describes a wide class of insulating, topologically non-trivial, solid-state materials~\cite{hasan_colloquium_2010,hasan_three-dimensional_2011,bernevig_topological_2013}.   
These ``topological insulators'' have now been observed in a variety of weakly correlated two- and three-dimensional  narrow gap semiconductors. 
More recently, the focus has shifted due to the discovery of gapless topological semimetals that are realized in the limit of a vanishing Dirac mass that results in the valence and conduction bands touching at isolated points in the Brillouin zone. 
While graphene~\cite{castro_neto_electronic_2009} is a commonly known two-dimensional Dirac semimetal, it was only recently that three-dimensional Dirac semimetals where discovered in Cd$_3$As$_2$~\cite{Neupane-2014,borisenko_experimental_2014,liu_stable_2014} and Na$_3$Bi~\cite{liu_discovery_2014,Xu-2015} as well as by doping the insulators Bi$_{1-x}$Sb$_{x}$~\cite{lenoir_transport_1996,ghosal_diamagnetism_2007,teo_surface_2008}, BiTl(S$_{1-\delta}$Se$_{\delta}$)$_2$~\cite{xu_topological_2011,sato_unexpected_2011}, and (Bi$_{1-x}$In$_x$)$_2$Se$_3$~\cite{brahlek_topological-metal_2012,wu_sudden_2013}.
Breaking inversion or time-reversal symmetry lifts the two-fold degeneracy of the Dirac touching point converting a single Dirac point  into two Weyl points. 
Three-dimensional Weyl semimetals have been clearly identified in the weakly correlated semiconductors with broken inversion symmetry \cite{Huang-2015,weng_weyl_2015} TaAs, NbAs, TaP, and NbP~\cite{Huang-2015,Xu2-2015,Xu3-2015,xu_experimental_2015,lv_experimental_2015,zhang_electron_2017}. 
There are also signatures of   strongly correlated Kondo-Weyl semimetals~\cite{lai_weylkondo_2018,chang_parity-violating_2018} in Ce$_3$Bi$_4$Pd$_3$~\cite{dzsaber_giant_2019} and YbPtBi~\cite{guo_evidence_2018}. 
On the other hand, the observation of Weyl semimetals that break time-reversal symmetry are much more rare with Mn$_3$Sn~\cite{kuroda_evidence_2017} being one example despite there being a number of proposed candidate materials with the pyrochlore iridates $R_2$Ir$_2$O$_7$ being a prominent example~\cite{Wan-2011,witczak-krempa_correlated_2014,goswami_competing_2017}. 

A nodal excitation spectrum extends well beyond electronic semimetals to also include the neutral Bogolioubov-de Gennes (BDG) quasiparticles in superconductors with non-trivial gap symmetry~\cite{leggett_theoretical_1975,hu_midgap_1994,volovik_universe_2009,schnyder_topological_2015}. 
In particular, gap symmetries that induce line nodes in the superconducting gap, which intersect the Fermi surface of the normal state at a finite number of points induces isolated nodal points in the BDG energy bands. 
For example, in three-dimensions a $p + ip$ superconductor can have Weyl points in its BDG band structure that produce gapless thermal Majorana excitations \cite{senthil_quasiparticle_2000,goswami_topological_2013,Wilson-2017} that are the neutral analog of an electronic Weyl semimetal.

Weyl nodes of opposite chirality act as sources and sinks of Berry curvature, which endow they system with  non-trivial topological properties. 
In time-reversal broken Weyl semimetals this can produce a non-zero anomalous Hall effect~\cite{burkov_anomalous_2014}, whereas in inversion broken Weyl semimetals it can lead to a non-zero photo-galvanic effect~\cite{goswami_optical_2015,ma_direct_2017}
in addition to a large non-linear response in transport \cite{wu_giant_2017,de_juan_quantized_2017}.
The topological nature of Weyl semimetals can also be revealed by the observation of their gapless surface states~\cite{Wan-2011} that exist along a line (i.e.\ an arc) connecting the projection of a pair of Weyl nodes of opposite helicity on the surface Brillouin zone (BZ). 
Topological Fermi arc surface states have now been observed on the surface of Weyl semimetals in angle resolved photo emission and scanning tunneling microscopy experiments~\cite{Xu3-2015,lv_experimental_2015,inoue_quasiparticle_2016,xu_observation_2016,yuan_quasiparticle_2019}.

While Weyl and massless Dirac fermions may not exactly exist in high-energy physics, their realization in the low-energy limit of an electronic  band-structure has opened the possibility to observe exotic high-energy phenomena in solid-state materials. 
In particular, the axial anomaly~\cite{adler_axial-vector_1969,bell_pcac_1969,nielsen_adler-bell-jackiw_1983} has consequences when massless Dirac or Weyl fermions are placed in parallel electric and magnetic fields in condensed matter experiments. 
In the lowest Landau level, this anomaly can be understood intuitively through a charge pumping process between Weyl nodes of opposite chirality~\cite{nielsen_adler-bell-jackiw_1983}.
The observation of the axial (or chiral) anomaly  has been seen indirectly in a number of Dirac and Weyl semimetals via the observation of a negative magnetoresistance~\cite{kim_dirac_2013,liang_ultrahigh_2015,li_chiral_2016} for parallel electric and magnetic fields and a large (or even colossal) positive magnetoresistance when the fields are perpendicular
~\cite{son_chiral_2013,parameswaran_probing_2014,gorbar_chiral_2014,burkov_chiral_2014,goswami_axial_2015,cano_chiral_2017}.
These observations are particularly remarkable since the connection to high energy physics breaks down due to the existence of an underlying band structure with a bounded bandwidth and band-curvature effects~\cite{vazifeh_electromagnetic_2013}. 

Moving away from the idealized band-structure limit, the lack of a hard-gap does not mean that topological protection, if it persists, cannot be related to an energy-gap protection vis-\`a-vis disorder and interactions.
The following manuscript reviews the current theoretical understanding of the properties and stability of the Weyl semimetal phase in the presence of short-range disorder while focusing on the non-perturbative rare-region effects.
As materials, these systems inherently have disorder (e.g., impurity defects and vacancies);
the type of disorder (long- or short-ranged) depends on its source.
For instance, long-ranged disorder can originate from Coulomb impurities~\cite{galitski_statistics_2007,beidenkopf_spatial_2011,skinner_coulomb_2014} that locally dope the Weyl cone due to screening charge ``puddles''  that lead to a non-zero density of states and conductivity for any finite density of impurities.
However, in the following review, we will focus on quenched short-ranged disorder that arises due to dislocations, vacancies, and neutral impurities, focusing on and highlighting rare-region effects. 
As we will see, this is a subtle non-perturbative problem in statistical physics and criticality, therefore interaction effects will not be discussed in this review (apart mean-field assumptions leading to the formation of superconducting BDG quasiparticles).

\begin{figure}
    \includegraphics[width=\textwidth]{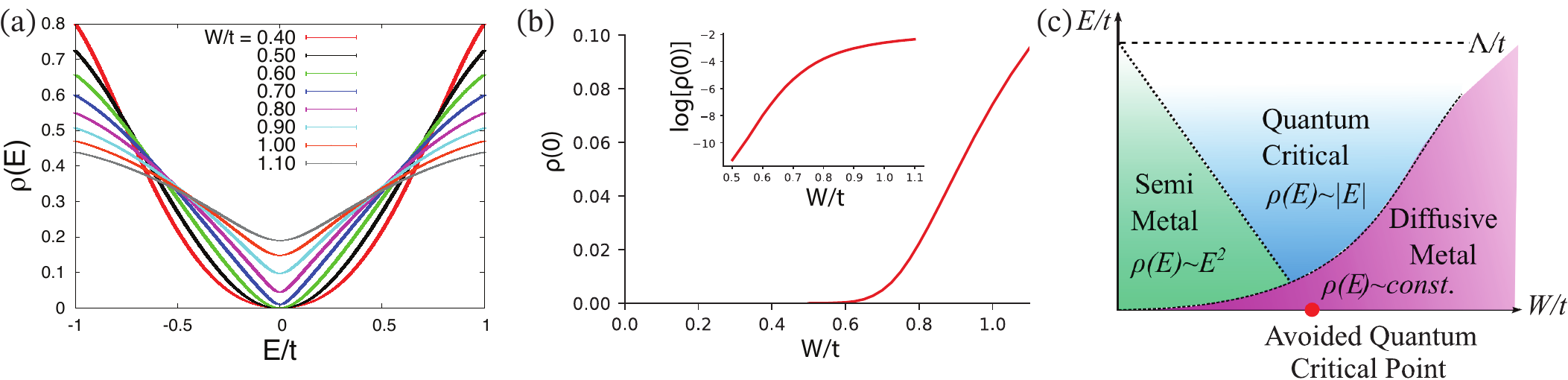}
	\caption{{\bf Evolution of the disorder averaged density of states depicting a cross over due to an avoided quantum critical point}. (a) Twist averaged density of state $\rho(E)$ vs energy $E$ showing the putative transition that is rounded into a crossover. 
	(b) The zero energy density states on a linear scale appears to depict a phase transition but on a log-scale (inset) it is non-zero but exponentially small at weak disorder (data is computed at finite size $L=71$ and KPM expansion order $N_C=2048$, see Sec.~\ref{sec:numericsDOS} for clarification). (c) A summary schematic cross over diagram exemplifying the avoided transition, with each relevant scaling regime at weak disorder including a semimetal, a diffusive metal, and quantum critical scaling that is rounded out by rare-region effects. From Ref.~\cite{Pixley-2016}.
	}
	\label{fig:AQCP_crossoever}
\end{figure}

Theoretically, the problem of disordered Weyl semimetals dates back to the work of Fradkin in 1986~\cite{fradkin_critical_1986,fradkin_critical_1986-1} where he showed that the semimetal is \emph{perturbatively} stable to the inclusion of short-ranged disorder. 
This stability persists (perturbatively) up to a putative critical point  were the density of states at the Weyl node becomes non-zero, which thus acts as an order-parameter for the transition, see Fig.~\ref{fig:AQCP_crossoever}(a,b).
This semimetal-to-diffusive metal quantum phase transition  was more accurately captured through a perturbative renormalization group (RG) calculation in $d=2+\epsilon$ dimensions~\cite{Goswami-2011}. 
Building upon this, Refs.~\cite{Leo-2015,Sergey-2015} generalized the nature of the transition to show how it arises in arbitrary dimensions by allowing the nodal touching points to have an arbitrary power law. 
Unlike Anderson localization~\cite{Evers-2008} (which occurs in these models at a larger disorder strength), the primary indicator is the density of states~\cite{fradkin_critical_1986}.
This perturbative and field theory picture has since been refined \cite{Leo-2015,Sergey2-2015,altland_effective_2015,altland_theory_2016,pixley_disorder-driven_2016,Bitan-2016,roy_universal_2016,syzranov_multifractality_2016,Louvet-2016,sbierski_weyl_2016,louvet_new_2017,luo_quantum_2018,luo_unconventional_2018,Balag-2018,roy_global_2018,brillaux_multifractality_2019,sbierski_strong_2019,sbierski_non-anderson_2020} with a better understanding for the theory at the critical point (though, as Ref.~\cite{sbierski_non-anderson_2020} points out, this understanding is still in question and ripe for further investigation, particularly with regards to the correlation length exponent).
Numerous numerical simulations  have found reasonable agreement with the perturbative prediction of a quantum critical point with close to the expected dynamic critical exponent~\cite{kobayashi_density_2014,Brouwer-2014,sbierski_quantum_2015,pixley_anderson_2015,pixley_disorder-driven_2016,liu_effect_2016,bera_dirty_2016,sbierski_weyl_2016,luo_quantum_2018,kobayashi_ballistic_2020}. 
However, whether or not the density of states remains truly non-zero in the supposedly stable semimetal phase, as in Fig.~\ref{fig:AQCP_crossoever}(b), was out of reach in early studies, where a large finite size effect appears in the density of states at the Weyl node energy. 

The perturbative picture of a stable semimetal phase was challenged in Ref.~\cite{nandkishore_rare_2014}, which demonstrated that rare region effects could induce a non-zero density of states at the Weyl node and make the system diffusive. 
Moreover, these rare regions originate from uncharacteristically strong disorder strengths (statistically ``rare'') producing a low-energy, quasi-bound wavefunction (i.e.\ falls off in a power law-fashion). 
However, initial numerical studies were not able to locate the rare states nor their effect on the low-energy density of states.
In two-dimensional Dirac materials, on the other hand, similar rare-resonances are reasonably well understood~\cite{wehling_dirac_2014}
and have been seen via scanning tunneling microscopy on vacancies on the surface of graphite~\cite{ugeda_missing_2010} and the $d$-wave high-temperature superconductor Bi$_2$Sr$_2$Ca(Cu$_{1-x}$Zn$_x$)$_2$O$_{8+\delta}$~\cite{pan_imaging_2000}. 
However, in two-dimensions, disorder is a marginally relevant perturbation in the RG sense (see Sec.~\ref{sec:perturbative_transition}) and therefore, the rare resonances are a sub-leading effect on average. 
In contrast, in three-dimensions the irrelevance of disorder implies the low-energy theory should be dominated by rare-resonances. 
We remark in passing that these rare states are the gapless analog of Anderson localized Lifshitz states~\cite{van_mieghem_theory_1992} that are well known to randomly fill in and soften the spectral band gap. 
The instanton calculus~\cite{suslov_density_1994,Leo-2015,yaida_instanton_2016} that is discussed below is directly borrowed from the  literature on Lifshitz states.

A systematic procedure to isolate and study rare eigenstates in disordered three-dimensional Weyl materials was put forth in Ref.~\cite{Pixley-2016}, which 
found excellent agreement with the  predicted form of the non-perturbative, quasilocalized eigenstates. 
It was also shown numerically that the density of states remains an \emph{analytic} function of both disorder and energy at the Weyl node; the purported perturbative transition is rounded into a crossover in the thermodynamic limit due to a finite length scale induced by rare regions. 
Nonetheless, as already observed in a number of numerical studies the predicted quantum critical scaling appears over a finite energy window between $E^*<E<\Lambda$ where $E^*$ is the rare region cross-over length scale that is non-zero in the presence of disorder and $\Lambda$ is a high-energy cut-off where the linear approximation of the nodal crossing points breaks down, see Fig.~\ref{fig:AQCP_crossoever}(c). 
Thus, the rare-region induced crossover was dubbed an avoided quantum critical point (AQCP). 
Naturally, these non-perturbative rare region effects could also destabilize the exotic topological properties of Weyl semimetals, something we will consider in detail in Sec.~\ref{sec:RR_topology}. 
As the theoretical description  of the perturbative quantum critical point and the rare region dominated limit are effectively expanding about two distinct ``mean field'' ground state wavefunctions the precise microscopic derivation of an effective theory for the AQCP has yet to be obtained. 
Ref.~\cite{Guararie-2017} has put forth a phenomenological field theory description of the AQCP in terms  of a gas of instantons interacting with a power law interaction that links the finite density of states with a finite-correlation length rounding out the putative transition. 

More recently, Buchhold \emph{et al.}~\cite{buchhold_vanishing_2018,buchhold_nodal_2018} challenged the scenario of a AQCP using a fluctuation analysis of the instanton field theory that found an exact zero for the density of states in the semimetal phase. 
Such an analysis focused on a single, linear, Weyl cone in the continuum and the previous numerics that found an AQCP, strictly speaking, did not apply. 
Following this, we put forth a numerical study of the nature of the avoidance in a single Weyl cone in the continuum~\cite{wilson_avoided_2020-1}. 
The numerical results found a strong avoidance, inconsistent with a stable semimetal phase. 

The purpose of this review is to present an exposition on the current understanding of rare region effects in disordered Weyl semimetals. 
Our goal is to present the most straightforward approach to find rare regions in numerical simulations as well as collect and discuss the various results that exist on the problem across a  range of microscopic models that all indicate that the perturbative QCP is rounded out into an avoided transition. 
The review therefore does not cover in depth the large body of RG literature that has developed around the perturbative transition and instead points the interested reader to the relevant literature as needed.
There are many open questions left to be explored, and we hope this review will aid other researchers to find, diagnose, and study non-perturbative rare region effects in semimetals so that they may resolve some of these open issues.
Our focus will be on the impact of non-perturbative rare region effects on destabilizing the Weyl semimetal (or superconductor) phase (Sec.~\ref{sec:DOS}), the nature of the transport and low-energy quasiparticle excitations (Sec.~\ref{sec:excitations_and_transport}),  the stability of the topological Fermi arc surface states (Sec.~\ref{sec:Fermi-arc}), and the persistence of the charge pumping process due to the axial anomaly (Sec.~\ref{sec:axial_anomaly}. 
New results on a random hopping model will be presented as additional evidence that particle hole symmetry cannot help stabilize the Weyl semimetal phase in the presence of disorder in Sec.~\ref{sec:ph_symmetry}.

\section{Model and set up}

In the following, we will be concerned with a broad class of models that can be described as
\begin{equation}
    H=H_{\mathrm{Weyl}}+H_{\mathrm{disorder}}
    \label{eqn:Ham-main}
\end{equation}
where $H_{\mathrm{Weyl}}$ describes the band structure that hosts linear touching points at low energy and $H_{\mathrm{disorder}}$ contains random, short-ranged disorder. 
Near some momenta $\mathbf K_W$, the band structure we are interested in has nodal linear touching points
\begin{equation}
    H_{\mathrm{Weyl}} \approx \sum_{\mathbf{K}_W}\sum_{{\bf k}} \psi_{{\bf k}}^{\dag}\left (v ({\bf k}-{\bf K}_W)\cdot\bm \sigma \right) \psi_{{\bf k}}
    \label{eqn:Weyl-form}
\end{equation}
where $v$ is the velocity of the Weyl cone, $\bm \sigma$ is a vector of the Pauli matrices, and $\psi_{\bf k}$ is a two-component spinor of annihilation operators. 
The resulting low-energy dispersion is given by $E({\bf k}) \approx \pm v|{\bf k}|$.
However, this approximation cannot hold across the entire Brillouin zone of the lattice.
First, for both simplified and DFT derived models of Weyl semimetals, there can be significant band curvature away from the band touching.
Even at the band-touching point, isotropy is not guaranteed, and furthermore, the Weyl cone can ``tilt'' leading to Type-II Weyl semimetals~\cite{Guararie-2017} with a finite Fermi surface at the nodal energy. Since Type-II Weyl semimetals have a finite density of states at the nodal energy, they will not host the disorder physics which we are reviewing in this paper.
And finally, by the Fermion doubling theorem~\cite{Nielsen81PLB,Nielsen81NPB1,Nielsen81NPB2} every lattice model will have an even number of Weyl cones.
In Sec.~\ref{sec:single_Weyl_node}, we discuss how we can exactly isolate a single node numerically and observe the same physics.

One of the main observables that we will focus on is the low energy density of states, which is defined as
 \begin{equation}
     \rho(E)=\frac{1}{L^3}\sum_i \delta(E-E_i)
     \label{eqn:dos}
 \end{equation}
where $E_i$ are the low energy eigenvalues and $L$ is the linear system size. 
A focus of this review is understanding the stability of the Weyl semimetal phase to disorder. 
One aspect of this  question boils down to whether or not the density of states at the Weyl node remains precisely zero in the presence of weak disorder, as scaling theory would predict~\cite{fradkin_critical_1986,fradkin_critical_1986-1,Goswami-2011}.
In the clean limit, for a three-dimensional Dirac or Weyl semimetal the scaling of the low-energy density of states goes like
\begin{equation}
    \rho(E)\sim \frac{1}{v^3}E^2,
    \label{eqn:doslowE}
\end{equation}
and central to this review is how $\rho(0)$ is modified in the presence of disorder. 
 
The particular form the of $H_\mathrm{disorder}$ depends on the physical problem, but one of the most studied problems is $\emph{potential}$ disorder.
In this case,
\begin{equation}
    H_{\mathrm{disorder}} = \sum_{\bf r} \psi_{\bf r}^\dagger V(\bf r) \psi_{\bf r},
\end{equation}
where the potential $V(\mathbf r)$ is short-range correlated $\braket{V(\mathbf r)V(\mathbf r')}_\mathrm{dis} = f(|\mathbf r - \mathbf r'|)$ and averages to zero, where $f(r) \ll  O(e^{-r/\xi})$ for $r\gg \xi$ and some $\xi$.
In lattice models, we will often take $\braket{V(\mathbf r)V(\mathbf r')}_\mathrm{dis} = W^2 \delta_{\mathbf r \mathbf r'}$.

In the absence of particle hole symmetry, it is important to remove the leading perturbative finite size effect.
For a random potential $V({\bf r})$ drawn from a normal distribution with zero mean and standard deviation $W$,
the leading perturbative correction to the energy is $ E_1=L^{-3}\sum_{{\bf r}}V({\bf r})\sim W/L^{3/2}$(random sign) that produces a random broadening to the energy levels. To eliminate this finite size effect, the potential can be shifted so that each sample has a random potential that sums exactly to zero. We denote the shifted by potential as $\tilde V({\bf r})=V({\bf r})-L^{-3}\sum_{{\bf r}}V({\bf r})$.

\subsection{An inversion symmetry broken Weyl semimetal}
 \label{sec:inversionWeyl}
The simplest model to consider only has one parameter in the bandstructure (the hopping) that gives rise to 
three-dimensional Weyl fermions with time reversal symmetry on a simple cubic lattice.
We will use this and the model in the following section to introduce and describe the numerical results, then later discuss other models and their differences.
This
model is defined as
\begin{equation}
H_{\mathrm{Weyl}}= \sum_{{\bf r},{\mu}={x},{y},{z}}\left( i t_{\mu}\psi_{{\bf r}}^{\dag}\sigma_{\mu} \psi_{{\bf r}+\hat{\mu}} + \mathrm{H.c}\right),
\label{eqn:weylTR}
\end{equation}
where $\psi_r$ is a two-component Pauli spinor, the $\sigma_{\mu}$ are the Pauli matrices with $\mu=x,y,z$. The clean dispersion for $t_x=t_y=t_z=t/2$ is given by 
\begin{equation}
    E_0(\bk) = \pm t \sqrt{ \sin(k_{x})^2+\sin(k_{y})^2+\sin(k_{z})^2},
\end{equation} 
and the model has eight Dirac points at the time reversal invariant momenta in the Brillioun zone.  
The band structure has time reversal symmetry but lacks inversion. 
In the presence of potential disorder, this model belongs to the Gaussian Orthogonal Ensemble (GOE) of random matrices.
This is due to an explicit symmetry in the model described by
\begin{equation}
    \psi_{\mathbf r} \rightarrow (-1)^{x+z}\sigma_y \psi_{\mathbf r}. \label{eq:symmetry}
\end{equation}
Despite the model having the anti-unitary symmetry $i\sigma_y K$ (where $K$ is complex conjugation in the real space basis), this model does not belong to a symmetry class with $T^2=-1$. 
To understand why, once we project this model into the subspaces described by the symmetry \eqref{eq:symmetry}, we obtain the Hamiltonian
\begin{equation}
    \tilde H_\mathrm{Weyl} = - \sum_{\mathbf r} [t_x(-1)^{x+z} c_\mathbf{r}^\dagger c_{\mathbf{r} + \hat{\mathbf x}}+t_y(-1)^{x+z} c_\mathbf{r}^\dagger c_{\mathbf{r} + \hat{\mathbf y}} + t_z c_\mathbf{r}^\dagger c_{\mathbf{r} + \hat{\mathbf z}}],
\end{equation}
and $\tilde H_\mathrm{dis} = \sum_\mathbf{r} c_\mathbf{r}^\dagger V(\mathbf r) c_\mathbf{r}$ where $c_{\mathbf{r}}$ is no longer a spinor.
This Hamiltonian makes it clear that the time-reversal operator for this model has $T^2=+1$.
Moreover, the time reversal symmetry can be removed by putting twisted boundary conditions on each sample of size $L^3$, accommodating the twist with a gauge transformation amounts to $t_{\mu} = t \exp(i \theta_{\mu}/L)/2$ with $-\pi <\theta_{\mu}<\pi$.  By taking a twist of $\bm \theta = (\pi,\pi,\pi)$ on even  $L$ we can maximize the finite size gap ($=2\sqrt{3}\sin(\pi/L)$) in our simulations. 

\subsection{A time-reversal symmetry broken Weyl semimetal}
\label{sec:TRbrokenWeyl}
  Another commonly studied model that we will discuss has broken time reversal symmetry that allows the Weyl points to occur at arbitrary points in the BZ. This model is defined as
  \begin{equation}
   H  =\sum_{\br,\hat{\nu}}\left(\psi_{{\bf r}}^{\dag}(t_{\nu}\sigma_z + t'_{\nu}\sigma_{\nu})\psi_{\br+\hat{\nu}} 
 + \mathrm{H.c} \right)
-\sum_{\br}\psi_{\br}^{\dag}(m\sigma_z)\psi_{{\bf r}}
\label{eqn:weylIS}
  \end{equation}
  where the hopping parameters are $t_{\nu}=t/2$ with $\nu=x,y,z$ and $t'_{\nu}=t'/2$ for $\nu=x,y$ and zero otherwise. The band structure is determined by the mass term $m$ and the hopping coefficients. 
  The dispersion of the model is given by
  \begin{equation}
      E_0({\bf k})=\pm \sqrt{t'^2(\sin(k_x)^2+\sin(k_y)^2)+[t\sum_{\nu} \cos(k_{\nu})-m]^2}.
  \end{equation}
  Setting $t'=t$, the band structure hosts a regime with 4 Weyl points for $|m|<t$ at ${\bf K}_W=(0,\pi,\pm K_2),(\pi,0,\pm K_2)$ where $K_2=\arccos(m/t)$, and 2 Weyl points for $t<|m|<3t$ at ${\bf K}_W=(0,0,\pm K_1)$ where $K_1=\arccos{m/t-2}$. At the transition between these  regions at $|m|=t,3t$ an anisotropic nodal dispersion appears that for $|m|=3t$ 
  goes like
  $E_0(\bk)\approx \pm t\sqrt{(k_z^2/2)^2+(k_x^2+k_y^2)}$ at low energy. The dispersion at $|m|=t$ is similar but there are 3 anisotropic nodal points at $\bK_N=(\pi,0,0),(0,\pi,0),(0,0,\pi)$. 
  Interestingly, the anisotropic nodes gives rise to a low energy density of states $\rho(E)\sim |E|^{3/2}$ and the renormalization group  (see Secs.~\ref{sec:RG} and~\ref{sec:beyond_linear_touching}) will also predict the existence of a putative semimetal-to-diffusive metal QCP and is an excellent case to compare with a Weyl node. 
  In the presence of potential disorder, this model  belongs to the GUE ensemble in the diffusive metal phase.
  
\subsection{Twisted boundary conditions}

At this point, it is useful to describe the theoretical and numerical implications of \emph{twisted boundary conditions}.
We will limit the discussion here to one-dimension for ease, but the discussion easily generalizes to higher dimensions and even beyond square lattices.

In the strictest sense, twisted boundary conditions for the single particle wave function in one-dimension (denoted $x$) of size $L$ identifies
\begin{equation}
    \phi(x+L) = \phi(x) e^{i\theta}
\end{equation}
with a twist $\theta$, which within our lattice model can be implemented via a gauge transformation
\begin{equation}
    \psi_x \rightarrow e^{-i\theta x/L}\psi_x,
\end{equation}
where $x$ is the coordinate over which the twisted boundary conditions are implemented, and we have restored periodic boundary conditions via this gauge transformation.
Physically, this is equivalent to threading a magnetic flux through the sample, breaking time-reversal symmetry.
However, there is an alternative way to understand these boundary conditions in terms of a time-reversal symmetric system.
If one considers an infinite system which is periodic with period $L$, then the Hamiltonian $H = \sum_{x,x'} \psi_{x'}^\dagger t_{x',x} \psi_x $, where we assume $t_{x',x}=0$ when $x'-x\geq L/2$, can be block diagonalized using Bloch's theorem, and the result is
\begin{equation}
    H_k = \sum_{x,x'=0}^{L-1} e^{-ik(x-x')} \psi_{x'}^\dagger t_{x',x} \psi_x,
\end{equation}
where $k\in [0,2\pi/L)$. 
If we identify $k=\theta/L$, then this is precisely the gauge-transformed Hamiltonian with twisted boundary conditions.
As such, the use of twisted boundary conditions can be equivalently viewed as randomly sampling points in the Brillouin zone of an \emph{infinite} system with a supercell of size $L$.

\section{Perturbative Transition}\label{sec:perturbative_transition}

Through the use of the self-consistent Born approximation and large-$N$ techniques, one can identify both the perturbative stability of the semimetallic phase and the existence of a purported transition~\cite{fradkin_critical_1986,fradkin_critical_1986-1}.
This method, however, does not give the correct critical theory (it is only correct in a large-$N$ approximation, see Ref.~\cite{syzranov_high-dimensional_2018} for a full discussion), and more sophisticated techniques are needed~\cite{Goswami-2011}.
We nonetheless review it here since it gives one of the simplest means to understand the perturbative stability and the existence of criticality.

\subsection{Self-consistent Born approximation}
Near a Weyl node, the first-quantized Hamiltonian takes the form
\begin{equation}
    H = -i v\bm \sigma \cdot \bm \nabla + V(\mathbf r), \label{eq:Weyl_first_quantized}
\end{equation}
where $V(\mathbf r)$ is a disorder potential with $\braket{V(\mathbf r)}_\mathrm{dis} = 0$ and $\braket{V(\mathbf r)V(\mathbf r')}_\mathrm{dis} = \Omega \delta(\mathbf r-\mathbf r')$ (regularized by a momentum space cutoff $\Lambda$ which is often supplied by lattice regularization).

The disorder-averaged Green's function can be written in terms of a self-energy $\Sigma$ as
\begin{equation}
    G(\mathbf k,\omega) = \frac{1}{\omega - v \mathbf k \cdot \bm \sigma  - \Sigma(\mathbf k,\omega) } = \left\langle\frac{1}{\omega - v \mathbf k \cdot \bm \sigma - V(\mathbf r)}\right\rangle_\mathrm{dis}.
\end{equation}
At second-order in $V(\mathbf r)$ and self-consistently, we can write 
\begin{equation}
    \Sigma(\omega) = \Omega \int_{|\mathbf k'|<\Lambda} \frac{d^3 k'}{(2\pi)^3} \frac{\omega-\Sigma( \omega)}{(\omega-\Sigma(\omega))^2 - |\mathbf k'|^2}.
\end{equation}
At $\omega=0$ this equation has an obvious solution of $\Sigma(0)=0$ which describes the stability of the semimetal to disorder, but as we increase $\Omega$, an imaginary solution $\Sigma(0)=i\gamma$ appears when 
\begin{equation}
 \Omega>\Omega_c=1/\int_{|\mathbf k'|<\Lambda} \frac{d^3 k}{(2\pi)^3} 1/ |\mathbf k'|^2,   
\end{equation}
indicating a finite density of states at the nodal energy and a diffusive metal.
This is also our first indication of a critical point $\Omega_c$.

This is perhaps the simplest way to construct the self-consistent Born approximation for this problem.
For more detailed accounts see Refs.~\cite{Ominato-2014,sinner_corrections_2017,Klier-2019}.

\subsection{Renormalization group and scaling theory}
\label{sec:RG}

With a one-loop renormalization group analysis \cite{Goswami-2011} one can identify a transition and obtain its critical properties.
Up to order $O(\Omega^2)$ using a $d=2+\epsilon$ expansion one finds the renormalization flow equations (after converting to the dimensionless $\Omega\rightarrow \Omega \Lambda/(2\pi^2 v^2)$) 
\begin{equation}
\begin{aligned}
    \frac{d\Omega}{d\log(\Lambda)} & = (2-d)\Omega+2\Omega^2,
    \\
    \frac{d v}{d\log(\Lambda)} & = v(z-1-\Omega).
    \end{aligned}
    \label{eqn:RG}
\end{equation}
In particular, in three-dimensions ($d=3$) the unstable fixed point is perturbatively accessible and is located at $\Omega_c=\epsilon/2$ with a dynamic exponent $z=1+\epsilon/2$ and correlation length exponent $\nu=1/\epsilon$. Taking $\epsilon=1$ (as appropriate for $d=3$) these analytic predictions give  $z=3/2$ and  $\nu=1$.
While existing numerical work finds agreement with the one-loop prediction for the dynamical exponent, the correlation length exponent has been harder to pin down\footnote{
We suspect that this is due to past studies using the density of states at the Weyl node or the conductivity as an indicator of the transition, but this is problematic as they are both expected to always be non-zero due to rare region effects.  Instead, the use of the second derivative of the density of states provides an unbiased indicator as discussed in Sec.~\ref{sec:numericsDOS}.
}.

In the absence of non-perturbative effects, scaling theory can be applied to extract the critical exponents from data on the density of states~\cite{kobayashi_density_2014}. The dynamical exponent can be found by the scaling of the density of states away from the nodal energy (at $\Omega=\Omega_c$)
\begin{equation}
    \rho(E,\Omega_c) \sim |E|^{d/z-1}
\end{equation}
while the scaling of $\rho(0)$ and its $2n$ derivatives can give us both critical exponents~\cite{kobayashi_density_2014,pixley_uncovering_2016}
\begin{equation}
    |\rho^{(2n)}(0)| \sim |\delta|^{-(z(2n+1)-d)\nu}, 
      \label{eqn:crit_zeroE}
\end{equation}
where $\delta=(\Omega-\Omega_c)/\Omega_c$ parametrizes the distance from the putative critical point. 
The critical scaling form in energy is then  
 \begin{equation}
     \rho(E,\Omega)\sim |\delta|^{\nu(3-z)}f_{\pm}(E|\delta|^{-\nu z}),
      \label{eqn:crit_finiteE}
 \end{equation} 
where $f_{\pm}(x)$ are scaling functions for positive and negative $\delta$. 
This analysis will be useful in Sec.~\ref{sec:numericsDOS} after we find and suppress rare region effects to reveal   the critical behavior in these models before the avoidance rounds it out.

Field theory descriptions and RG calculations of the transition in various forms have gone well beyond what have detailed here. 
As the current review focuses on the \emph{avoidance} of the transition and we will not cover them in this review but point the reader to the relevant literature in Ref.~\cite{syzranov_high-dimensional_2018}.

\section{Rare resonances in semimetals}
\label{sec:rare_resonances}
To capture rare regions in Dirac and Weyl semimetals, a natural starting point is to treat them as sufficiently dilute so that we can ignore there interaction and replace each rare region with a strong potential well. Such an approach works well in describing similar rare resonances in two-dimensional Dirac semimetals \cite{wehling_dirac_2014}. 
The quintessential way in which each rare region is then described is via an exact solution to Eq.~\eqref{eq:Weyl_first_quantized} with $V(\mathbf r) = V_0 \Theta(r - R)$ \cite{nandkishore_rare_2014}, a spherical potential of strength $V_0$ and radius $R$, replacing the rare region with a simple but strong potential.
There are solutions for this potential at all energies $E$; first, with spherical symmetry total angular momentum is a good quantum number, so we label eigenstates with radial quantum number $n$, total angular momentum $j$, and $J_z$ quantum number $j_z$ such that
\begin{equation}
\psi_{E; n,j,j_z}(r,\theta,\phi) = f_{E,j}(r) \chi^+_{j,j_z}(\theta,\phi) + i g_{E,j}(r) \chi^-_{j,j_z}(\theta,\phi)
\end{equation}
the spinors $\chi^\pm_{j,j_z}$ are eigenstates of $ \bm \sigma \cdot \mathbf L$ with total angular momentum quantum numbers previously defined and orbital angular momenta $\ell^\pm = j \mp \tfrac12$.
The full solution of these equations is in Refs.~\cite{nandkishore_rare_2014,buchhold_vanishing_2018,buchhold_nodal_2018} but for our purposes it is important to note that when $E=0$, bound-state solutions exist for particular values (when $j=1/2$, then $V_0 R = p\pi$ with integer $p$) and has 
\begin{equation}
    f_{0,j}(r) \sim r^{-(j+3/2)},  \label{eq:rarestate}
\end{equation}
for $r\gg R$ and $g_{0,j}(r>R)=0$.
These solutions still exist for modifications of the potential away from a square well, and when the condition for their existence is not quite satisfied (when slightly off-resonance), there is still an effect on the density of states (though if it effects the density of states exactly at the nodal energy is disputed Refs.~\cite{buchhold_vanishing_2018,buchhold_nodal_2018,pires_randomness_2020} at the time of this writing).

\subsection{Numerical evidence of rare states}

  \begin{figure}[t!]
	\centering
 \includegraphics[width=\textwidth]{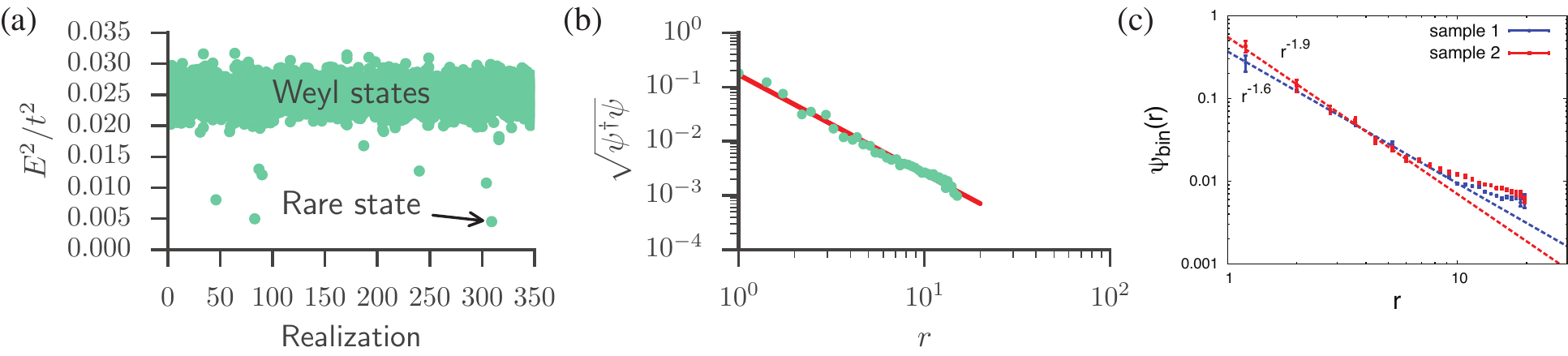}
	\caption{{\bf Rare eigenstates and a simple way to find them}. (a) A demonstration of finding rare eigenstates by studying the distribution of the square of low-energy eigenvalues computed with Lanczos on $H^2$ as a function of the disorder realization for the Hamiltonian $H$ in Eq.~\eqref{eqn:weylIS} at weak disorder $W=0.5t$ and system size $L=18$. The perturbative Weyl states form a dense band of states that make up a peak in the density of states (as shown in Fig.~\ref{fig:diracpeaks}) whereas the rare states occur at lower energy below the peak filling in the finite size gap (induced by the twist). (b) The corresponding rare eigenstate state labelled in (a) displaying its power law decay from its maximal value with a power law $|\psi({\bf r})|\sim 1/r^{1.83}$ in good agreement with the theoretical expectation in Eq.~\eqref{eqn:rarestate}. (c) Two distinct rare states found at low energy and weak disorder for $W=0.6t$, $L=25$, with power law exponents $=1.6$ (blue circles) and $1.9$ (red squares) in the model in Eq.~\eqref{eqn:weylTR}. (a) and (b) are from Ref.~\cite{Wilson-2018} and (c) is from Ref.~\cite{Pixley-2016}.}
	\label{fig:rareWFs}
\end{figure}
 
At weak disorder and small $L$ we can loosely break spectrum into two parts, perturbatively dressed Weyl states and non-perturbative rare states. 
This separation has been argued to be parametrically enhanced in lower dimensional models with a modified power law dispersion \cite{Garttner-2015}.

\begin{figure}[t!]
	\includegraphics[width=\textwidth]{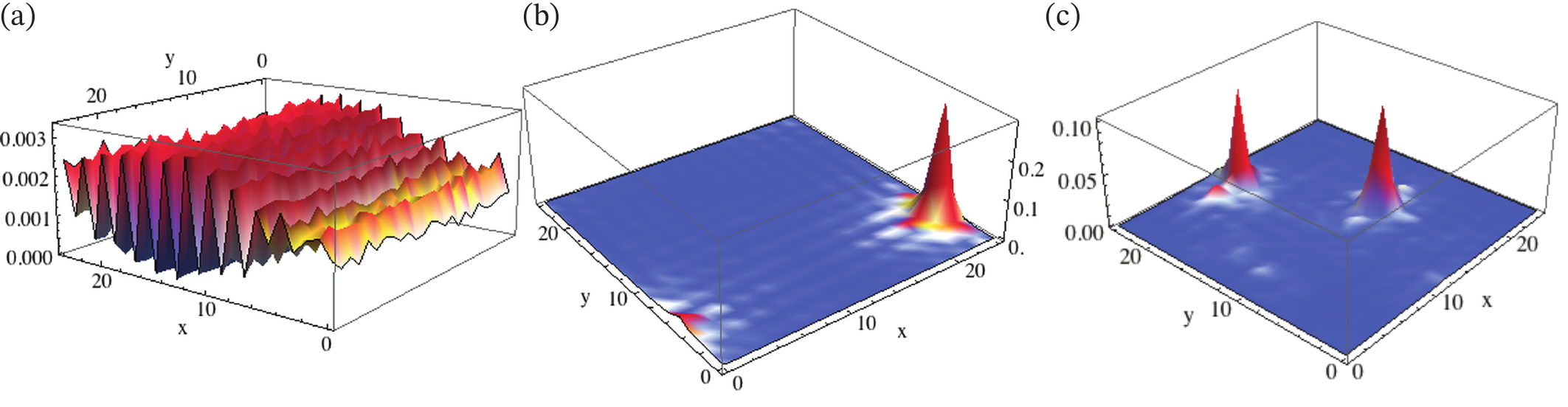}
	\caption{{\bf Eigenstates at weak disorder viewed through  the projected probability density $\sum_z |\psi_E(x,y,z)|^2$ for the model in Eq.~\eqref{eqn:weylTR}.} (a) An example of a perturbative state at low energy $E$ taken from the first Weyl peak at $W=0.3t$ and $L=25$. The state is delocalized across the system and the twist of $\theta_x = \pi/2$ produces a standing wave in the $x$-direction. 
	(b) A rare state obtained for a disorder strength $W=0.5t$ and a (c) bilocalized rare state with two rare regions for disorder strength $W=0.66t$, both are values of $W$ are below the avoided transition. Figures reproduced form Ref.~\cite{Pixley-2016}.
	}
	\label{fig:WFs}
\end{figure}

The irrelevance of disorder implies that at weak disorder and finite $L$, the majority of the low energy eigenstates will be perturbatively renormalized Weyl states. 
In particular, applying perturbation theory yields perturbed energies that go like $E\sim 1/L$ (i.e.\ they still disperse linearly in momentum) with a broadening at weak disorder like $\sim W^2/L^2$. In addition, the perturbed plane 
 wave states
 \begin{equation}
     \psi^{\mathrm{pert}}_{\bk,\pm}(\br)=\psi^{0}_{\bk,\pm}(\br)+\psi^{1}_{\bk,\pm}(\br)+\dots,
 \end{equation} 
 acquire an $L$ independent contribution of random sign and magnitude $W$, 
 \begin{eqnarray}
    \psi^0_{\bk,\pm}&=&\frac{1}{L^{3/2}}e^{i \bk \cdot \br}\phi_{\bk,\pm},
    \\
     \psi^{1}_{\bk,\pm}(\br) &\sim& W \mathrm{(random \,\,\,\, sign)},
 \end{eqnarray}
 where $\phi_{\bk,\pm}$ is a normalized two-component spinor as shown in Ref.~\cite{Pixley-2016}. 
 As shown in Fig.~\ref{fig:WFs}(a), this is in good agreement with numerical results at weak disorder.

In the model represented by Eq.~\eqref{eqn:weylTR}, Weyl points exist at the time-reversal symmetric points in the Brillouin zone.
Thus for even system sizes $L$, we can choose a twist $\bm \theta = (\pi,\pi,\pi)$ to move the Weyl states maximally away in energy.
In general, with careful analysis one can identify where the Weyl points are in the mini-Brillouin zone defined by the twisted boundary conditions and move them as far from zero energy as possible with a clever combination of system size and twists $\bm \theta$.
This is useful for identifying states that develop due to a rare resonance: they should be weakly modified via a twist in the boundary conditions.
In fact we can see in Fig.~\ref{fig:rareWFs}(a) an example of this where the typical (plane-wave-like) states all live in a narrow band (of width $W^2/L^2$) around the mean energy, and rare states are filling in the finite-size gap induced by the twisted boundary conditions. 
As we predicted by Eq.~\eqref{eq:rarestate}, we expect these states to be quasibound, and in all likelihood they come from the smallest total angular momentum $j=1/2$, implying~\cite{nandkishore_rare_2014} 
\begin{equation}
     \psi^{\mathrm{rare}}(\br)\sim \frac{1}{|\br-\br_0|^2},
     \label{eqn:rarestate}
\end{equation}
about some point $\mathbf r_0$.
Indeed, this is the case, such as in Fig.~\ref{fig:WFs}(b).
The numerically observed rare states across the models in Eqs.~\eqref{eqn:weylTR} and \eqref{eqn:weylIS} all show a power law decay, Fig.~\ref{fig:rareWFs}(b,c), with power law values ranging from $\sim 1.5-2.0$, in good agreement with the analytic expectation. 
Further, they originate from a disorder potential $V(\mathbf r_0)$ much larger than its surrounding values.

Another clear distinction between perturbative and rare states can be seen in ``twist dispersions'' (i.e.\ determining how the energy eigenvalues vary under a twist). 
The perturbatively dressed states have Weyl cones in the dispersion with a reduced velocity $v-v_0\sim  -W^2$. 
The rare states on the other hand strongly reconstruct the band structure and they  disperse very weakly away from the crossing points. 
It is also possible to find samples with multiple rare regions in a single sample. 
For example, as shown in Ref.~\cite{Pixley-2016} and Fig.~\ref{fig:WFs}(c), two rare states produces a pair of quasilocalized states that have a non-zero tunneling matrix element between them.

\section{Density of states}
\label{sec:DOS}
In the following section we focus on the effects of rare regions on the low energy density of states. First, we present a derivation of the density of states using instanton calculus and then its numerical estimate showing it remains non-zero for any disorder strength. This brings us to a measure of the avoidance, as well as a phenomenological theory for the AQCP and  a systematic way to tune the avoidance length scale.

\subsection{Instanton calculation of density of states}

The issue of density of states was raised within the first paper on the subject, Ref.~\cite{fradkin_critical_1986}.
Formally, one can write the density state for the Weyl equation, Eq.~\eqref{eq:Weyl_first_quantized} setting $v=1$, as 
\begin{equation}
    \rho_V(E) = \frac1{L^3}\sum_n \delta(E - E_n)
\end{equation}
where $(-i\bm \sigma \cdot \nabla + V(\mathbf r))\psi_n = E_n \psi_n$. 
This can be written as a functional integral
\begin{equation}
    \rho_V(E) = \frac{1}{L^3} \int \mathcal D[\psi(\mathbf r), \chi(\mathbf r), \Upsilon]e^{i\int d^3 r \chi^\dagger(\mathbf r) (E + i\bm \sigma \cdot \nabla - V(\mathbf r)) \psi(\mathbf r) + i \Upsilon[(\int  d^3 r \psi^\dagger(\mathbf r) \psi(\mathbf r)) - 1]} .
\end{equation}
Determining the disorder averaged quantity $\rho(E)$ then for a Gaussian random field with $\braket{V(\mathbf r)V(\mathbf r')} = W^2 K(\mathbf r-\mathbf r')$ can be accomplished also with a functional integral
\begin{equation}
    \rho(E) = \int \mathcal D[V] e^{-\frac{1}{2W^2} \int d^3 r \, d^3 r' V(\mathbf r) K^{-1}(\mathbf r - \mathbf r') V(\mathbf r')} \rho_V(E),
\end{equation}
where $K^{-1}$ is defined such that $\int d^3 y' K^{-1}(\mathbf y- \mathbf y') K(\mathbf y'-\mathbf y'') = \delta^3(\mathbf y-\mathbf y'')$.
Taking the saddle point approximation of this averaged quantity at $E=0$ then amounts to solving the non-linear integro-differential equation
\begin{equation}
    \left[-i \bm \sigma \cdot \nabla - \chi_0 W \int d^3 \mathbf r' K(\mathbf r-\mathbf r')\psi^\dagger(\mathbf r') \psi(\mathbf r') \right]\psi(\mathbf r) = 0, \label{eq:instanton_eqn}
\end{equation}
for normalized solutions (with $\chi_0$ allowed to be freely chosen).
As a mean-field, this will lead to a change in the density of states 
\begin{equation}
    \delta \rho(0) = \frac{1}{L^3} e^{-\frac{\chi_0^2}2 \int d^3 r \, d^3 r' \, \psi^\dagger(\mathbf r) \psi(\mathbf r) K(\mathbf r - \mathbf r') \psi^\dagger(\mathbf r') \psi(\mathbf r')}.
\end{equation}
As it turns out, there are solutions to \eqref{eq:instanton_eqn} that resemble the rare resonances previously discussed in Sec.~\ref{sec:rare_resonances} (see Ref.~\cite{nandkishore_rare_2014} for the full analysis). 
Within some suitable approximations, given some correlation length of the disorder $\xi$ and strength $W$, the change of the density of states for a \emph{single} rare resonance will follow $\delta \rho(0) \sim \frac{1}{L^3} e^{-C \frac{\xi}{2W^2}}$ for some constant $C$ while if there are a finite density $\rho_0$ of these, we expect
\begin{equation}
    \delta \rho(0) \sim \rho_0 e^{-C \frac{\xi}{2W^2}},\label{eq:theory_rare_rho0}
\end{equation}
for some constant $C$, at the level of the saddle-point approximation for a \emph{single} rare state.
This picture is complicated by work that includes fluctuations on top of the mean-field (in Refs.~\cite{buchhold_vanishing_2018,buchhold_nodal_2018}) and finds an extensive number of zero-modes which drive $\delta \rho(0) \rightarrow 0$ exactly.
However, we present the mean-field analysis here since it fits the numerical data remarkably well and leave the resolution of the fluctuation calculation with numerics as an open problem at present in the field.

\begin{figure}[t!]
	\centering
		\includegraphics[width=0.49\textwidth]{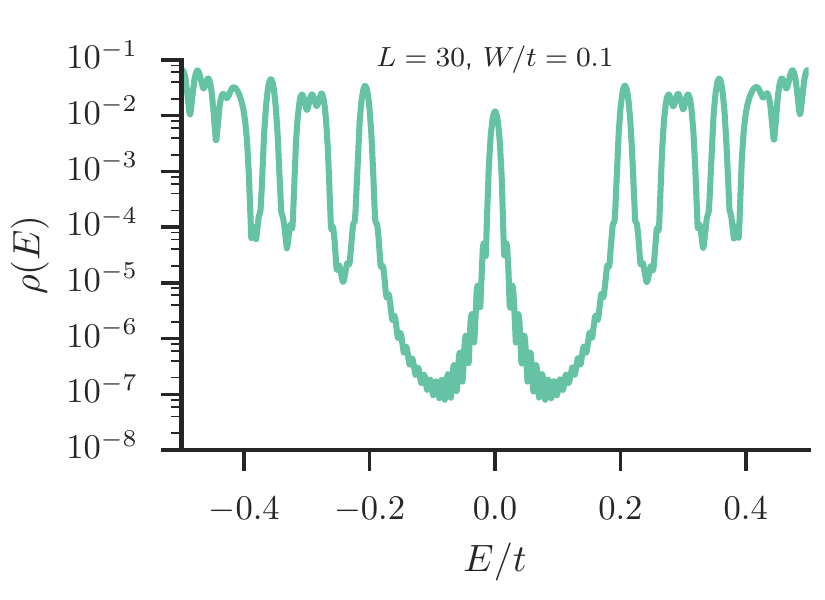} 
		\includegraphics[width=0.49\textwidth]{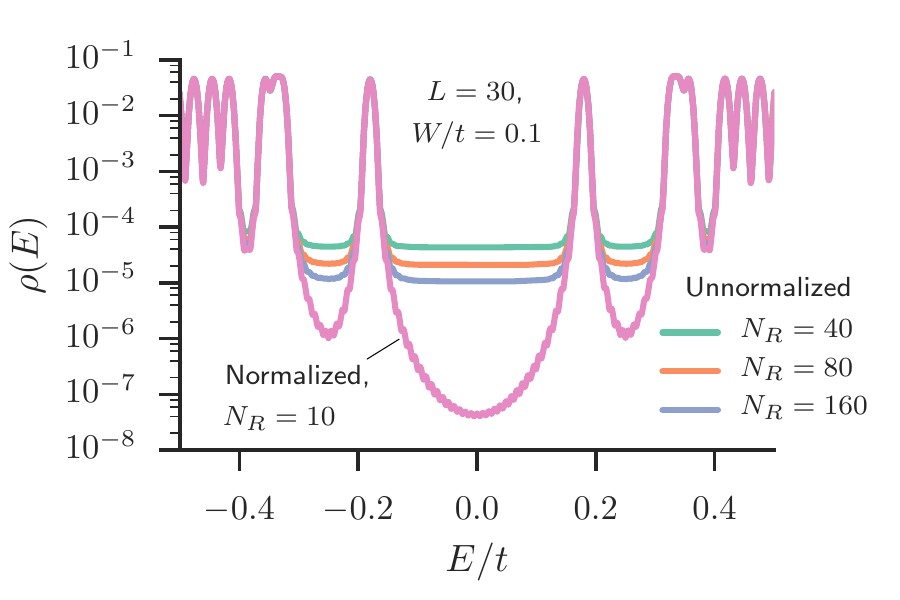} 		
	\caption{{\bf Density of states $\rho(E)$ as a function of energy $E$ displaying low-energy perturbative states}. (Left) The use of periodic boundary conditions (and certain system sizes) can place a zero energy Weyl state in the spectrum that becomes broadened by disorder. This produces a large finite size effect in the zero energy density of states that obscures the rare region contribution that is orders of magnitude smaller at weak disorder. (Right) Using a twist to move the Dirac state to a non-zero energy  introduces a finite size gap of order $\sim 1/L$. The low energy peaks are refereed to as ``perturbative Weyl peaks'' that are composed of Weyl eigenstates that have been perturbatively renormalized by the disorder. The use of normalized random vectors (or random phase vectors) allows for an accurate determination of the zero energy density of states. From Ref.~\cite{Wilson-2017}. 
	}
	\label{fig:diracpeaks}
\end{figure}

 \subsection{Non-zero, analytic, density of states at the Weyl node}
 \label{sec:numericsDOS}

 The following results are obtained using numerically exact approaches that utilize sparse matrix vector multiplication to reach large system sizes beyond that accessible within exact diagonalization. 
In particular, the density of states (DOS) is computed using the kernel polynomial method (KPM) that expands the density of states using Chebychev polynomials to a finite order $N_C$ \cite{Weisse-2006}. 
The calculation of the DOS using KPM involves a stochastic trace, which to resolve the exponentially small rare region contribution needs to be done as accurately as possible. 
This can be achieved by evaluating the stochastic trace using either random phase vectors~\cite{alben_exact_1975,iitaka_random_2004} or random vectors with a fixed normalization~\cite{hams_fast_2000} (see Fig.~\ref{fig:diracpeaks}), either of which reduces the noise in the signal by orders of magnitude allowing for the rare region contribution to be properly estimated well below the avoided quantum critical point.

\begin{figure}[t!]
	\centering
		\includegraphics[width=0.48\columnwidth]{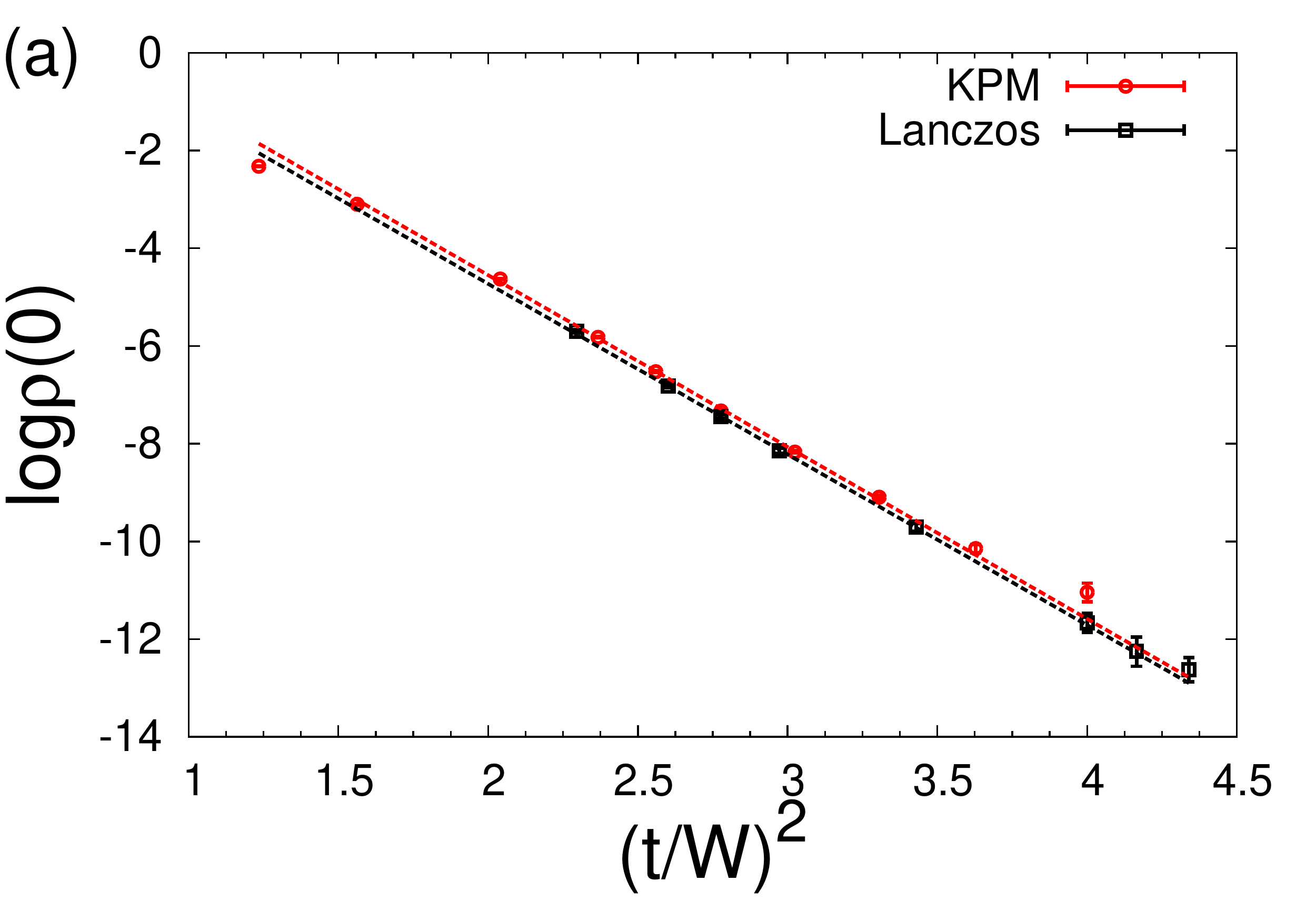}
	\includegraphics[width=0.48\columnwidth]{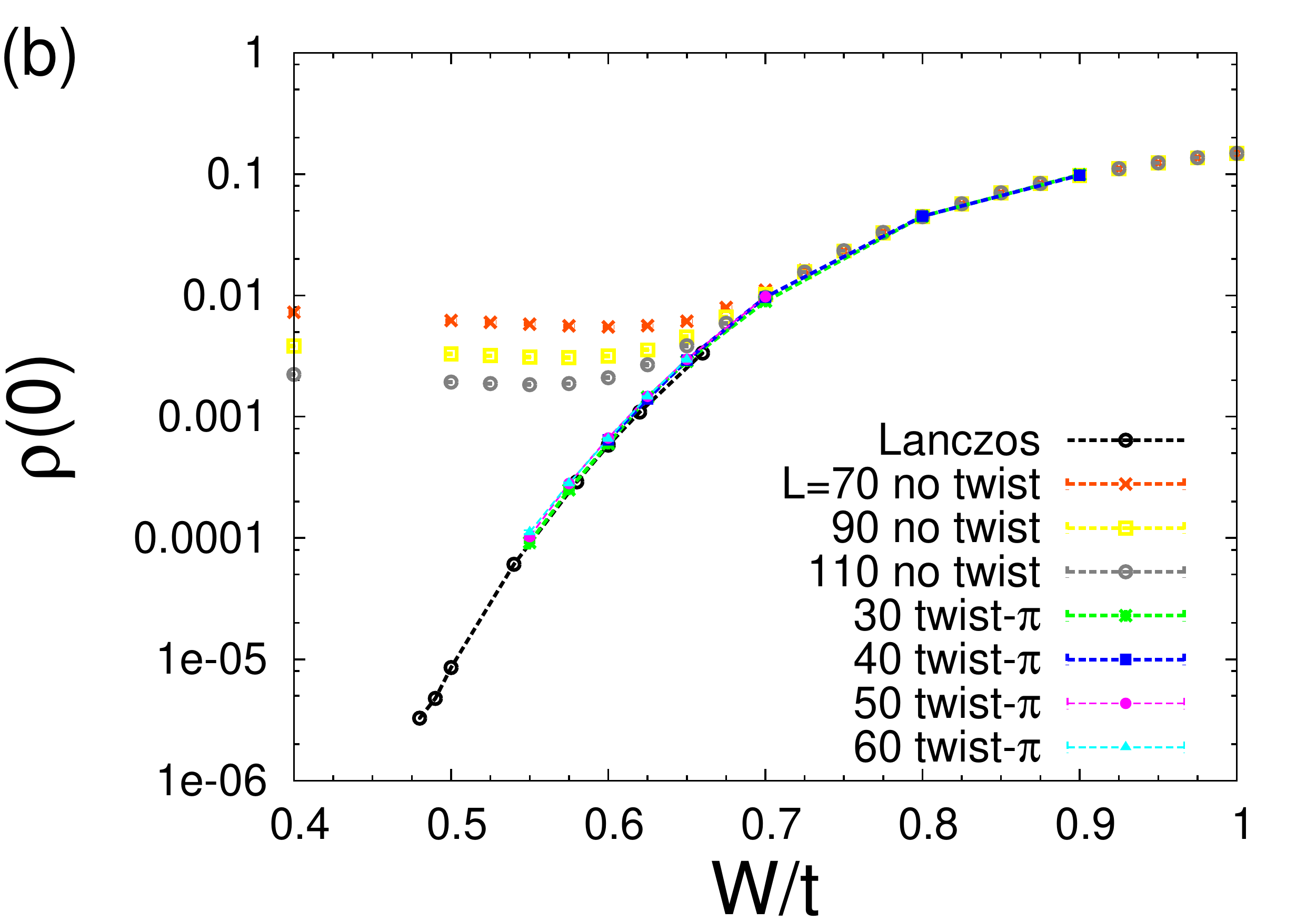}
	\caption{{\bf Rare region contribution to the density of states in the model in Eq.~\eqref{eqn:weylTR}}. (a) The log of the zero energy density of states computed using the KPM and Lanczos (at a small but finite energy) averaged over a large number of samples at fixed twist values displaying an excellent fit to the rare region form in  Eq.~\eqref{eqn:rareDOS} for over four orders of magnitude. (b) Comparing the zero energy density of states with periodic boundary conditions and even $L$ that produces a large finite size effect due to perturbative Dirac states (see Fig.~\ref{fig:DOS}). This finite size effect masks the rare region contribution that is converged in system size deep into the semimetal regime (below the AQCP) and several orders of magnitude below the Weyl peak artifacts . Instead the density of states is always non-zero, though exponentially small at weak disorder. Taken from Ref.~\cite{Pixley-2016}.}
	\label{fig:DOS}
\end{figure}
 
Focusing on the model in Eq.~\eqref{eqn:weylTR} with a twist of $\bm{\theta} = (\pi,\pi,\pi)$ (to push the Weyl states as far away from $E=0$ as possible similar to Fig.~\ref{fig:diracpeaks}). 
The relevant question is now whether rare states fill in the finite size gap in an extensive manner that is $L$-independent. 
The average density of states in the semimetal regime has a rare region contribution that fills in the finite size gap and is $L$-independent. 
At higher energies the Weyl peaks round into Weyl shoulders that are still spaced like $\sim 1/L$. 
Extracting the $L$-independent contribution of the low energy density of states is shown in Fig.~\ref{fig:DOS} and compared with the results of periodic boundary conditions that host a clean Dirac state at zero energy. 
The rare eigenstates thus contribute a non-zero but exponentially small DOS 
 \begin{equation}
     \log \rho(0) \sim -(t/W)^2
     \label{eqn:rareDOS}
 \end{equation}
 in good agreement with the theoretical prediction of Ref.~\cite{nandkishore_rare_2014} and Eq.~\eqref{eq:theory_rare_rho0}.

The presence of a non-zero DOS at the Weyl node suggests that the semimetal-to-diffusive metal QCP has been rounded out into a crossover. 
In order to test for the existence of a disorder driven quantum critical point, it is essential to determine if the density of states remains an analytic function of energy.
The analytic properties of the density of states can be determined by directly evaluating its energy derivatives. 
As the transition is avoided, in the presence of randomness, the DOS in the vicinity of the Weyl node energy remains analytic, namely
\begin{equation}
    \rho(E) = \rho(0) + \frac{1}{2!}\rho''(0)E^2 + \frac{1}{4!}\rho^{(4)}(0)E^4 + \dots.
    \label{eqn:dos-analytic}
\end{equation}
While the coefficients of this expansion can in principle be estimated using a fit there is a weak amount of rounding introduced by the fit that is not intrinsic to the problem. 
Therefore, it is more accurate to  use the KPM expansion of $\rho(E)$ to obtain the exact expression for the derivatives $\rho''(0)$ and $\rho^{(4)}(0)$  in terms of a similar Chebyshev expansion, the explicit expression for $\rho''(0)$ can be found in Ref.~\cite{pixley_uncovering_2016}.

The analytic nature of the DOS according to Eq.~\eqref{eqn:dos-analytic} is computed with KPM and shown in Fig.~\ref{fig:tuning-avoidance}(a), demonstrating that $\rho''(0)$ has a broad maximum for increasing disorder strength.
The location of the peak provides an accurate and unbiased estimate of the avoided QCP. 
We stress that the DOS at zero energy cannot be used as an indicator of the transition as it is \emph{always} non-zero, 
and we suspect that a similar issue plagues the DC conductivity, though a careful numerical analysis of the rare region contribution to transport remains to be done.
Focusing on $\rho''(0)$, we saturate the height of the peak as we increase the system size and KPM expansion order as shown in Fig.~\ref{fig:tuning-avoidance}. 
As  $\rho''(0)$ does not diverge, we conclude that the DOS remains an analytic function of energy, and the perturbative transition is rounded out into a cross over. 

\subsection{A theory for the avoided transition}
Motivated by the success of the instanton saddle-point point approximation capturing the rare region contribution to the density of states Gurarie in Ref.~\cite{Guararie-2017} treated the system at weak disorder as a gas of instantons interacting with a pair potential in a power-law like form (due to the quasi-localized rare states) that goes as $\sim a_0^{4}/r^2$ where $a_0$ is the characteristic size of a rare state. The resulting effective field theory produces a single particle Green's function that is analytic with a finite correlation length that is given by
\begin{equation}
    \xi \sim 1/\rho(0)\sim \mathcal{A} e^{a (t/W)^2},
\end{equation}
here $\mathcal{A}$ and $a$ are non-universal constants. This result confirms that the non-zero density of states at the Weyl node produces a finite length scale, albeit exponentially large at weak disorder, which rounds out the perturbative transition. As the value of the density of states and its dependence on the disorder strength depends strongly on the type of disorder distribution chosen, it is an interesting to understand how to tune the avoidance length scale $\xi$, which we now turn to.

 \subsection{Tuning the strength of the avoidance}
 
\begin{figure}[t!]
\includegraphics[width=\textwidth]{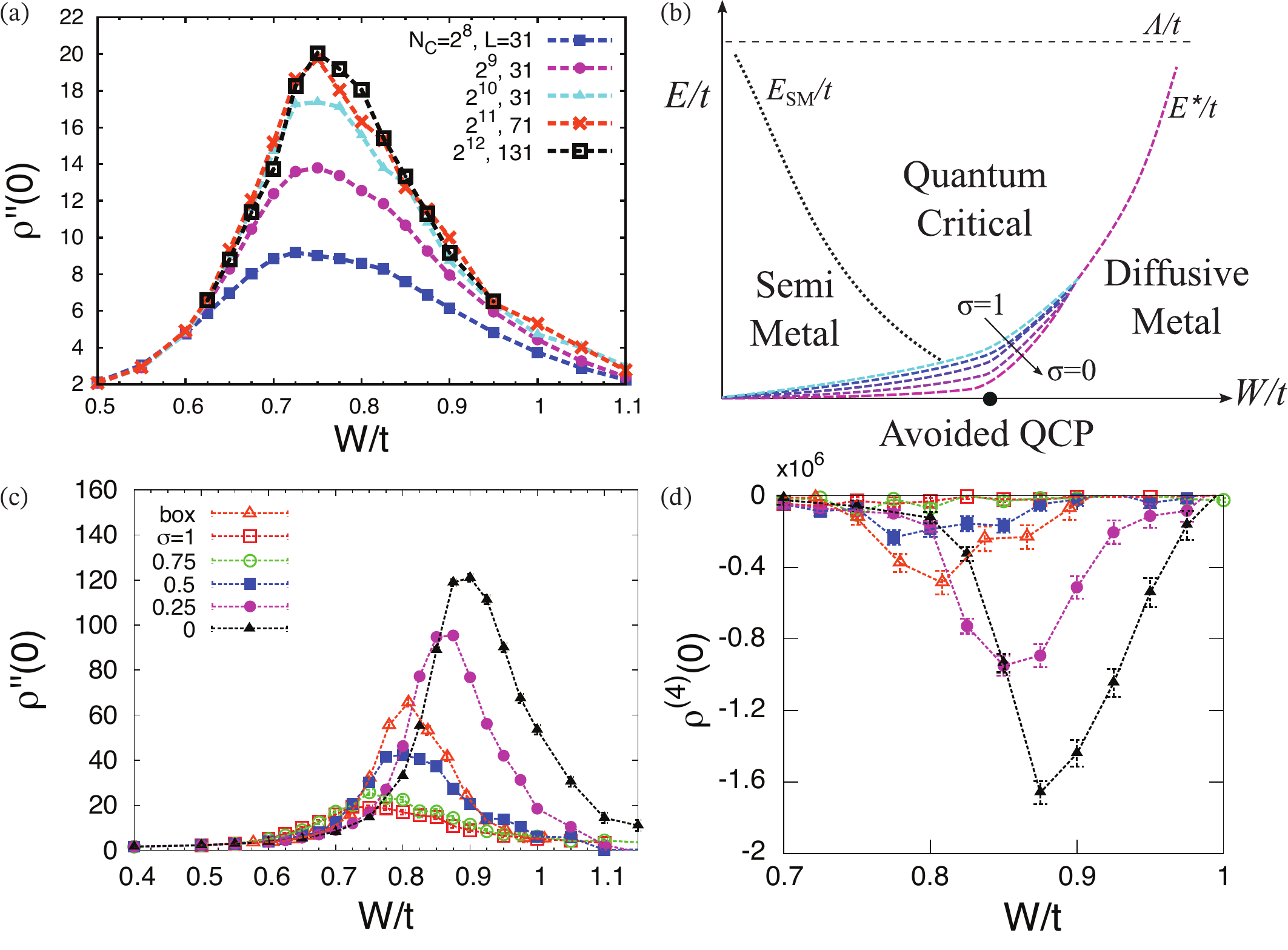}
	\caption{{\bf The strength of avoidance}. 
	(a) Converging the peak in $\rho''(0)$ for the case of Gaussian disorder with increasing system size and KPM expansion order, which demonstrates that the density of states remains an analytic function of disorder and energy. 
	(b) The strength of the avoidance can be tuned by varying the tails of the probability distribution; here we use a double Gaussian distribution with mean $\pm W \sqrt{1-\sigma^2}$ and a standard deviation $\sigma W$ so that tuning $\sigma$ controls the tails of the two Gaussians. 
	As $\sigma \rightarrow 0$ we approach a binary distribution (i.e.\ randomly sampling $\pm W$) where the avoidance is the weakest. 
	(c) $\rho''(0)$ as a function of disorder for various values of $\sigma$, demonstrating it tunes the strength of the avoidance.
	We also compare to the standard box distribution uniformly sampled between $[-W/2,W/2]$ which has a much weaker avoidance than Gaussian but not as sharp as binary disorder. 
	(d) The fourth derivative showing the avoidance can also be tuned in the higher derivatives of the low-energy density of states. 
	Taken from Ref.~\cite{pixley_uncovering_2016}.}
	\label{fig:tuning-avoidance}
\end{figure}

 In order to access the properties of the AQCP we need to first remove the rounding of the transition by suppressing rare events and obtain an accurate estimate of the location of the avoided transition (which is in fact difficult as it is a cross-over).
 First,
 it is possible to tune the probability to generate rare events rather naturally through varying the disorder distribution $P[V]$. As shown in Ref.~\cite{pixley_uncovering_2016}, tuning the tails of $P[V]$ allows us to control the strength of the avoidance as measured by the size of $\rho''(0)$. 
 It was found that binary disorder has a rather strong suppression of rare events and so interpolating between Gaussian and binary via a double Gaussian distribution is a rather nice choice of $P[V]$ to control the strength of the avoidance. 
 As shown in Fig.~\ref{fig:tuning-avoidance} (b), as we tune from Gaussian to binary $\rho''(0)$ goes from exhibiting a strong avoidance with a highly rounded peak and a small value  to significantly sharpen as rare events are suppressed.
 Importantly, we can use the location of the maximum of the peak in $\rho''(0)$ as an unbiased, accurate estimate of the avoided transition. We stress that any estimate of the location of the AQCP based on the density of states $\rho(0)$ or the conductivity at the Weyl node will be plagued by the fact that they are always non-zero due to rare regions and neither have a truly sharp change in the thermodynamic limit. However, there have been several attempts in the literature to estimate the location of the AQCP using where the density of states looks like it is lifting off of zero on a linear scale, which tends to greatly underestimate $W_c$ that leads to in inaccurate estimate of $\nu$.
 
\begin{figure}[t!]
	\centering
\includegraphics[width=\textwidth]{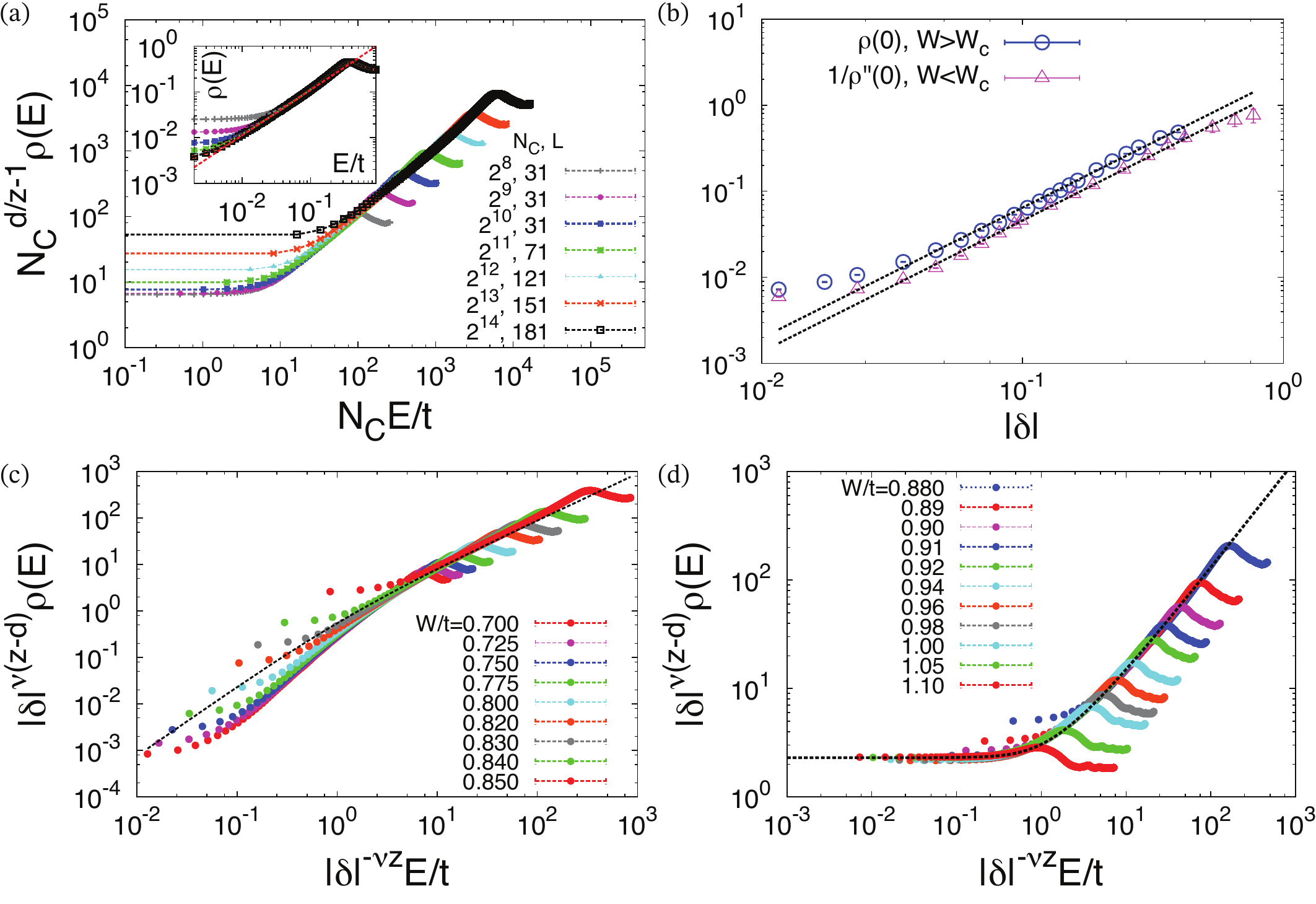}
	\caption{{\bf The perturbative critical scaling of the DOS is clearly revealed using binary disorder to suppress rare region effects for the model in Eq.~\eqref{eqn:weylTR}.} (a) Finite KPM expansion order scaling of the energy dependence of the DOS at the avoided transition with binary disorder utilizing the scaling form in Eq.~\eqref{eqn:crit_Nc}, which shows a good collapse. (Inset) Unscaled density of states showing the power-law scaling at the transition $\rho(E)\sim|E|^{d/z-1}$ with $z=1.50\pm 0.05$ for increasing $N_C$. (b) The power-law scaling of the zero energy density of states (and its second derivative) above (and below) the AQCP showing a power-law dependence with a fit to Eq.~\eqref{eqn:crit_zeroE} (dashed lines) that yields a correlation length $\nu=1.01\pm 0.06$ from $\rho(0)$ and $\nu=1.02\pm 0.08$ from $\rho''(0)$. Collapsing the data following the finite energy scaling form in Eq.~\eqref{eqn:crit_finiteE} is shown in (c) and (d) for $W<W_c$ and $W>W_c$ respectively. The dashed lines are the cross-over functions computed by integrating the RG equations and are explicitly presented in the Appendix in Ref.~\cite{pixley_disorder-driven_2016}. Results from Ref.~\cite{pixley_uncovering_2016}. }
	\label{fig:dos-critical}
\end{figure}

 In the limit of binary disorder we aptly posed to estimate the critical properties of the AQCP before it becomes rounded out by rare regions at the lowest energy scales. 
 Using binary disorder we estimate the critical exponents $\nu$ and $z$ from the data in Fig.~\ref{fig:dos-critical} using the critical scaling form in $d=3$ dimensions \cite{kobayashi_density_2014,pixley_uncovering_2016} of the zero energy density of states and its $2n$ derivatives as we described in Sec.~\ref{sec:perturbative_transition} (near the critical point we can replace $\Omega$ with $W$, though we note here $\Omega\propto W^2$).
 In addition, within the KPM the finite expansion order leads to an infrared energy scale that goes like $\delta E \sim 1/N_C$ and thus we can use $N_C$ scaling in the manner we do finite $E$ scaling to obtain similar scaling functions. 
 For example, if we consider the scaling of the density of states at the AQCP $W=W_c$ at sufficiently large $L$ we have 
 \begin{equation}
     \rho(E,W=W_c,N_C)\sim N_C^{1-d/z}g(E N_C)
     \label{eqn:crit_Nc}
 \end{equation}
 where $g(x)$ is an unknown scaling function.
 We stress that due to the avoided nature of the transition these scaling forms are only applicable in the energy regime $E\gg E^*$ where $E^*$ is non-zero energy scale which is induced by the rare regions of the random potential.
 
 The collapse of the numerical data, shown in Fig.~\ref{fig:dos-critical}(a), works over a full decade until the avoidance leads to deviations from the critical scaling form. 
 The power law nature of the zero-energy density of states and the second derivative above and below the transition respectively are shown in Fig.~\ref{fig:dos-critical}(b), which displays a critical power law as in Eq.~\eqref{eqn:crit_zeroE} over a full decade until the avoidance rounds it out at small $\delta$ due to the non-zero DOS. 
 In Figs.~\ref{fig:dos-critical}(c) and (d), we collapse the data in terms of the finite energy scaling forms in Eq.~\eqref{eqn:crit_finiteE} above and below the transition. The collapsed data is compared with the cross over functions that are obtained by integrating the RG equations in Eq.~\eqref{eqn:RG}, which display remarkable agreement (after adjusting the two bare RG energy scales); see Ref.~\cite{pixley_disorder-driven_2016} for the explicit cross-over functions.

 \section{Excitations and Transport}
 \label{sec:excitations_and_transport}
 We now turn to the nature of the single particle excitation spectrum and transport properties within linear response.
 
 \subsection{Quasiparticle excitation spectrum}
 \label{sec:qp-excitations}
 The excitation spectrum of the system can be extracted through the analytic structure of the single particle Green's function. 
 Averaging over disorder, the retarded Green's function --- expanded in the vicinity of single-particle poles --- takes the form
 \begin{equation}
     G_{\pm}(\bk, \omega) \approx \frac{Z_{\pm}(\bk)}{\omega - E_{\pm}(\bk)+i\gamma_{\pm}(\bk)}
 \end{equation}
 where $Z_{\pm}(\bk)$ is the quasiparticle residue, $E_{\pm}(\bk)$ is the single particle spectrum in the valence and conduction band basis (labelled $\pm$), and $\gamma_{\pm}(\bk)=1/\tau_{\pm}(\bk)$ is the damping rate or inverse quasiparticle lifetime.
 
 To get a sense of how rare regions can alter the excitation spectrum we  begin by  reviewing the $T$-matrix results from Ref.~\cite{pixley_single-particle_2017}. 
 If we work in the dilute impurity limit of a collection  of spherically symmetric square wells the potential takes the form $V(\br)=\sum_{j=1}^{N_{\mathrm{imp}}}\lambda_j \Theta(b -|\br - \bR_j|)$ for $N_{\mathrm{imp}}$ impurities  with a fixed width $b$ of random strength $\lambda_j$ taken from a Gaussian distribution $P[\lambda]$. The full Green function is obtained through Dyson's equation $G(\bk,\omega)_{\alpha\beta}^{-1}=G^{(0)}_{\alpha\beta}(\bk,\omega)_{\alpha\beta}^{-1}-\Sigma_{\alpha\beta}(\bk,\omega)$ where $G^{(0)}_{\alpha\beta}(\bk,\omega)$ is the free Green. In the dilute impurity limit ( i.e. $N_{\mathrm{imp}}/V \ll 1$ for a volume $V$)  the self-energy is momentum independent and given by
 \begin{equation}
     \Sigma_{\alpha\beta}(\omega)=\delta_{\alpha \beta}\frac{N_{\mathrm{imp}}}{V}\int d\lambda P[\lambda]T^{(\lambda)}(\bk,\bk)|_{v|\bk| = E}
 \end{equation}
 where $T^{(\lambda)}(\bk,\bk)$ denotes the $T$-matrix of for scattering off of a single impurity of strength $\lambda$ and $P[\lambda]$ denotes the normal distribution used to sample $\lambda_j$. 
 For each realization of $\lambda$, the $T$-matrix is computed analytically by solving the quantum mechanical scattering problem off a single well. The $T$-matrix is given by
 \begin{equation}
     T^{(\lambda)}(\bk,\bk)=-\frac{2\pi v}{k}\sum_j (2j+1) \frac{e^{2i\delta_j}-1}{2ik}
 \end{equation}
 and it is solely determined by the scattering phase shift $\delta_j$ (in the angular momentum channel $j$) that is known exactly for this problem (see Ref.~\cite{nandkishore_rare_2014} for the explicit expression). 
As a result, the leading long-wave length contribution arises from only considering the $j=1/2$ and 3/2 channels, and allow for a direct calculation of the single particle excitation properties $E_{\pm}(\bk),\gamma_{\pm}(\bk),Z_{\pm}(\bk)$ at both the perturbative and non-perturbative level from the simplified problem of a single impurity.
 \begin{figure}[t!]
	\centering
\includegraphics[width=\textwidth]{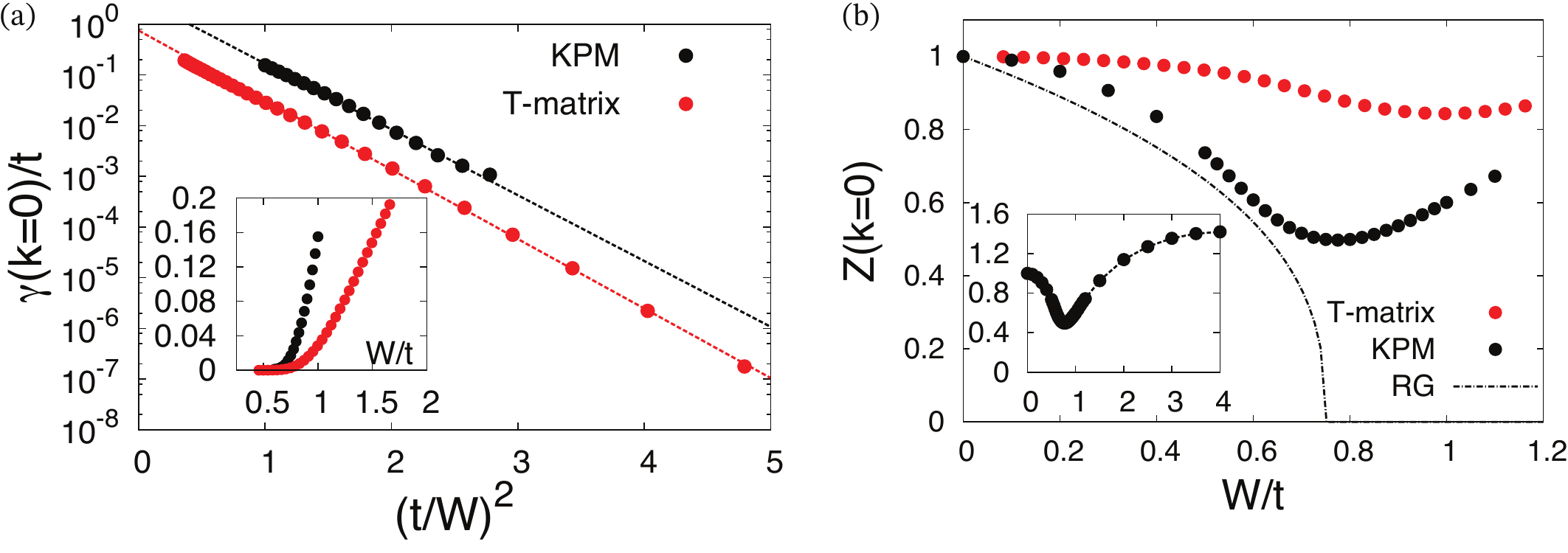}
	\caption{{\bf The nature of single-particle excitations at the Dirac node at} ${\bf k}=0$ {\bf for the model in Eq.~\eqref{eqn:weylTR}}. (a) The exponentially small damping at weak disorder following the non-perturbative rare region form in Eq.~\eqref{eqn:damping}. (b) The quasiparticle residue at the Dirac node remains finite across the AQCP as demonstrated in the exact numerics with Gaussian disorder that we compare with the RG that misses the rare-regions and the T-matrix which misses the perturbative transition. (Inset) The quasiparticle residue out to large disorder. Results taken from Ref.~\cite{pixley_single-particle_2017}.
	}
	\label{fig:damping}
\end{figure}
 
To develop a complete solution of the problem, in Ref.~\cite{pixley_single-particle_2017} the single particle properties where extracted by computing the Green's function with the KPM that was compared with the $T$-matrix approximation. 
At weak disorder at the Weyl node ($\bK_W$) the average low energy dispersion remains linear with momentum 
\begin{equation}
    E_{\pm}(\bk) = \pm v(W)|\bk-\bK_W|,
\end{equation}
where the Weyl velocity $v(W)$ is reduced (perturbatively like $-W^2$). On the other hand, the damping rate is exponentially small in disorder, namely for $\bk \approx \bK_W$
 \begin{equation}
 \begin{aligned}
    \gamma(\bk) & \approx \gamma(\bK_W) + B|\bk-\bK_W|^2
    \\
     \gamma(\bK_W) &= Ae^{-a(t/W)^2},
     \label{eqn:damping}
     \end{aligned}
 \end{equation}
which agrees well with a $T$-matrix approximation as shown in Fig.~\ref{fig:damping}(a).
While in this regime $Z(\bk)\approx \mathrm{const.}$ and $0<Z(\bk)<1$, the notion of Weyl quasiparticle excitation exists, but  they are no longer sharp due to the non-zero lifetime. 
Namely for sharp excitations to exist at the Weyl node requires $|E_{\pm}(\bk)| \gg \gamma(\bk)$ as $\bk \rightarrow \bK_W$ so that the (effectively) Lorentzian peaks in the spectral function at $E_{\pm}(\bk)$ have negligible overlap. 
However, due to rare regions we necessarily have $\gamma(\bK_W)\gg |E_{\pm}(\bk)|$ for $\bk \approx \bK_W$ as $|E_{\pm}(\bk)|\rightarrow 0$ in this regime destroying the notion of a sharp excitation in the vicinity of the Weyl point. 
As shown in Fig.~\ref{fig:damping}(b) the quasiparticle residue never vanishes implying that the Green function remains an analytic function at the Weyl node, consistent with an AQCP.

\begin{figure}[t!]
	\centering
\includegraphics[width=\textwidth]{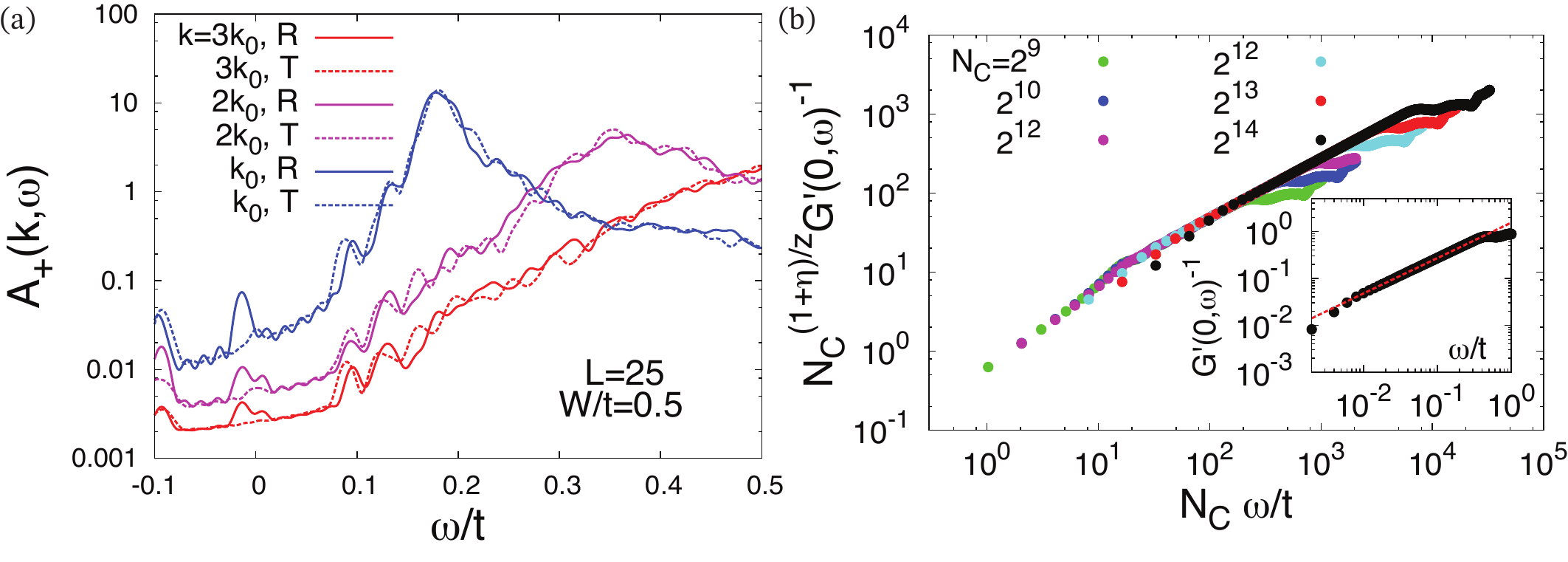}
	\caption{{\bf Dynamical scaling of single-particle excitations in momentum space for the model in Eq.~\eqref{eqn:weylTR}}. (a) The single-particle spectral function for individual samples at different momentum $k_0=2\pi/L$ comparing a typical (i.e.\ perturbative) sample and a rare sample with a non-perturbative state, which produces a non-zero overlap near $\omega=0$ for any momentum due to the power-law nature of the wavefunction. (b) The critical nature of the Green's function at the Dirac node ($\bk=0$) and binary disorder at the AQCP  with excellent data collapse in terms of the finite KPM scaling form in Eq.~\eqref{eqn:G_NC}. (Inset) The real part of the inverse Green's function at ${\bf k}=0$ at the AQCP showing a clear power-law scaling in agreement with Eq.~\eqref{eqn:G_crit} until it rounds out at the lowest energy due to the avoidance. From Ref.~\cite{pixley_single-particle_2017}.}
	\label{fig:spectral_func}
\end{figure}

To understand the  wavefunction overlap between a Weyl plane-wave state and a quasilocalized rare eigenstate at weak disorder, we study the off-shell energy and momentum dependence of the spectral function  as shown in Fig.~\ref{fig:spectral_func}(a). As the rare state is rather local in real space, it is broad in momentum space and should have a non-zero overlap with an arbitrary plane wave state. 
This is consistent with the numerical results, which show the rare sample produces finite spectral weight at low energy for each momentum considered, whereas a typical sample does not.
 
 The power law nature of these low energy excitations is renormalized at finite energy by the AQCP taking on the critical properties for a system with $z=3/2$ and $\nu=1$. In addition, the anomalous dimension  of the AQCP (denoted as $\eta$) is  defined via the Green function as 
 \begin{equation}
     \mathrm{Re}\,G(0,\omega)\sim \frac{1}{\omega^{(1+\eta)/z}}.
     \label{eqn:G_crit}
 \end{equation}
 where the power law only holds for frequencies above the rare region energy scale ($\omega \gg E^*)$. Numerical calculations using binary disorder to suppress rare regions have obtained $\eta=0.13\pm 0.04$, which is in good agreement with a modified RG scheme from Ref.~\cite{Goswami-2011} that yields $\eta=1/8$. As shown in Fig.~\ref{fig:spectral_func}(b) we can collapse the data using a finite KPM ($N_C$) scaling form 
 \begin{equation}
     \mathrm{Re}\,G(0,\omega=0,N_C)\sim N_C^{(1+\eta)/z} h(\omega N_C),
     \label{eqn:G_NC}
 \end{equation}
 where $h(x)$ is an unknown scaling function for well over two decades.
 
 \subsection{Transport at the Weyl node}
We close this section by discussing the diffusive transport properties of disordered Weyl semimetals that is mediated by the rare regions and will not review the work on the perturbative transition \cite{Sergey-2015,roy_universal_2016}. 
Currently, demonstrating these effects numerically is an open problem, despite the fact that hints of the avoided transition have been seen in scaling of the conductance~\cite{Brouwer-2014,sbierski_quantum_2015} and wavepacket dynamics~\cite{kobayashi_ballistic_2020}.
In particular, Ref.~\cite{Brouwer-2014} identified the perturbative semimetal regime and the diffusive metal phase in the average conductance, however no signs of rare region induced diffusion where found in the semimetal regime. Nonetheless, signs of critical scaling have been observed in the behavior of the average conductance as a function of disorder strength and Fermi energy~\cite{sbierski_quantum_2015} as well as a decreasing wavepacket velocity near the AQCP~\cite{kobayashi_ballistic_2020}.

\begin{figure}[t!]
	\centering
		\includegraphics[width=0.5\columnwidth]{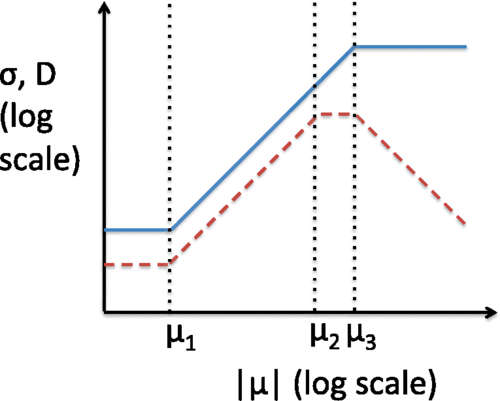} 
	\caption{{\bf The three distinct scaling regimes of  the conductivity $\sigma$ (solid blue line) and diffusion constant $D$ (red dashed line) as a function of the chemical potential $\mu$ due to rare regions.} At high energies $\mu>\mu_3 \sim v^{5/2}\rho(0)^{1/2}b^{1/2}/W$ the self consistent born described in Sec.~\ref{sec:perturbative_transition} works well. In the intermediate regime $v^2\rho(0)b\sim\mu_3<\mu<\mu_2\sim v^{3/2}\rho(0)^{1/2}$ and the low energy regime $\mu<\mu_1$ the density of states is dominated by rare regions. In the low energy regime transport is mediated by tunneling between rare regions [as in the bi-quasi-localized wavefunction in Fig.~\ref{fig:WFs} (c)] that produces a diffusion constant $D=l^2/\tau$ where $l$ is the typical hopping length and $\tau$ is the inverse damping at the Weyl node which is exponentially large at weak disorder. This approach misses the presence of the AQCP and so it is absent from these scaling regimes. 
	Reprinted figure with permission from \href{https://doi.org/10.1103/PhysRevB.89.245110}{Nandkishore \emph{et al.}~Phys.\ Rev.\ B 89, 245110 (2014)} \cite{nandkishore_rare_2014}. Copyright (2014) by the American Physical Society.}
	\label{fig:transport}
\end{figure}

As shown in Ref.~\cite{nandkishore_rare_2014} at sufficiently weak disorder there are three expected regimes in energy for the transport properties at linear response as seen in  
in Fig.~\ref{fig:transport}. 
At high energies  the behavior is dominated by the self-consistent Born approximation of Sec.~\ref{sec:perturbative_transition}. At the lowest and intermediate energy scales, tunneling between rare regions dominate the behavior leading to diffusive transport and a finite DC conductivity. Identifying these signatures in numerics remains a pressing open question. 

Ref.~\cite{Holder-2017} has built upon the previous $T$-matrix approach making it self consistent to compute the DC conductivity. 
Solving the self-consistent equations yields a general result for the DC conductivity for an arbitrary scattering rate $\gamma(\mu)$ at the chemical potential $\mu$, that is given by
\begin{equation}
    \sigma_{dc}(\mu)=\sigma_0\frac{\mu^2+\gamma(\mu)^2}{\gamma(\mu)},
\end{equation}
where $\sigma_0>0$ and depends on the microscopic details.
These results are also completely consistent with the behavior sketched in Fig.~\ref{fig:transport}.
At the Weyl node energy $\mu=0$ and the conductivity is determined by the scattering rate (and $\sigma_0$), which is exponentially small but non-zero at weak disorder due to rare regions as shown in Eq.~\eqref{eqn:damping} and Fig.~\ref{fig:damping}. Thus the DC conductivity is  expected be non-zero for any non-zero disorder strength and rare regions have converted the Weyl semimetal to a diffusive metal. In future work, it will be worthwhile to test these predictions numerically.

 \section{Rare regions and the topological properties of Weyl semimetals}
 \label{sec:RR_topology}
 
 The effects of rare regions on the topological properties of Dirac and Weyl semimetals remains a pressing question of ongoing research. Here, we briefly review work on the effects of rare regions on Fermi arc surface states and the axial anomaly. It will be very exciting to build on these results to understand the effects of rare regions on the nature of quantum oscillations in thin films and the negative magnetoresistance that has been used to infer the existence of the axial anomaly.
 
 \subsection{Fermi arc surface states}
\label{sec:Fermi-arc} 
 The Fermi arc surface states are robust to surface disorder \cite{shtanko_robustness_2018}.
 However, while not robust to bulk disorder, the surface states still survive to some extent with bulk disorder: perturbatively becoming power-law~\cite{gorbar_origin_2016} and dissolving into the bulk~\cite{slager_dissolution_2017} hybridizing with rare states~\cite{Wilson-2018}.
 Recent work \cite{brillaux_fermi_2020} further shows that a perturbative transition on the surface can occur, separate from the bulk (which rare events probably also make into an avoided criticality).
 We review here how the surface character persists despite these facts.
 
 \begin{figure}[t!]
	\centering
		\includegraphics[width=\textwidth]{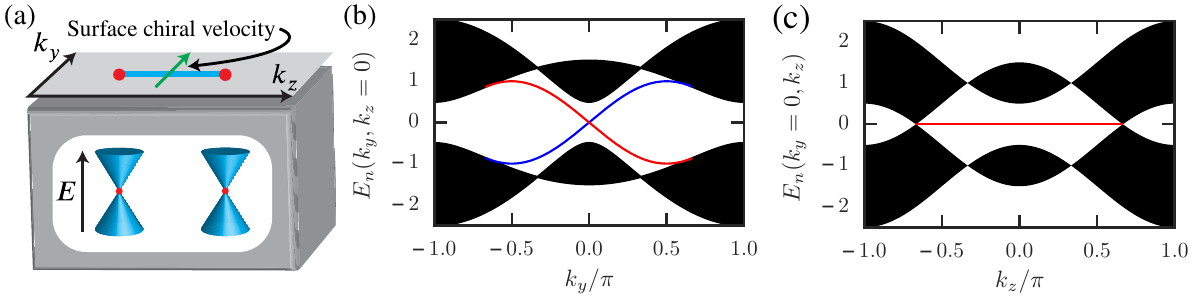}
	\caption{{\bf Fermi arc surface states in the model defined in Eq.~\eqref{eqn:weylIS} for two Weyl nodes with open boundary conditions along the $x$-direction.} (a) A schematic of the surface Brillouin zone displaying the projection of the Weyl points onto the surface and the Fermi arc of surface states connecting them. Their chiral velocity is denoted with a green arrow. 
	(b) and as a function of $k_z$ for $k_y=0$ (c) demonstrating the surface Fermi arc dispersion in Eq.~\eqref{eqn:surface_dispersion}. The surface states of the opposing surfaces are shown in red and blue and the bulk states are in black. Taken from Ref.~\cite{Wilson-2018}.}
	\label{fig:fermi_arcs}
\end{figure}

 The interplay of Fermi arc surface states and rare regions can be most easily studied in the model in Eq.~\eqref{eqn:weylIS} with two Weyl points and a single Fermi arc between them on particular surfaces. For the present model this represents a Fermi arc along a straight line between the projection of the two Weyl points onto either the $x-z$ or $y-z$ surface, see Fig.~\ref{fig:fermi_arcs} (a). In the clean limit of the model in Eq.~\eqref{eqn:weylIS} (setting $t=t'$) we obtained the exact surface eigenstates to be (for opening the boundary along $x$)
 \begin{eqnarray}
    \psi_S(x,y,z)&=&\frac{1}{L}e^{i(y k_y+zk_z)}f_S(x)\phi,
    \\
    f_S(x)&=& \sqrt{1-\lambda^2}\,\lambda^{x-1}
 \end{eqnarray}
 where the spinor $\phi^T= (1,-1)/\sqrt{2}$, we have defined 
 \begin{equation}
      \lambda(k_y,k_z)=m/t-\cos(k_y)-\cos(k_z),
 \end{equation}
 and  $\lambda \ge 0$ for the surface state solutions to exist.
 These surface state are exponentially bound to the surface decaying into the bulk like $\psi_S\sim e^{-x/\xi(\bk_{\perp})}$ where $\xi=1/\ln \lambda$ and $\bk_{\perp}=(k_y,k_z)$ is the momentum on the surface. The  dispersion of these Fermi arc surface states  is
 \begin{equation}
     E_S(\bk_{\perp})=t\sin(k_y)
     \label{eqn:surface_dispersion}
 \end{equation}
 and the opposing surface have states that disperse with the opposite chirality. The dispersion computed with open boundaries in the $x$-direction along $k_y$ (for $k_z=0$) and along $k_z$ (for $k_y=0$) are shown in Fig.~\ref{fig:fermi_arcs}(b) and (c), respectively, showing excellent agreement with Eq.~\eqref{eqn:surface_dispersion}. Even though these states only disperse along $k_y$, the results presented in this section are qualitatively insensitive to any non-zero curvature along the arc that will make them disperse along both $k_y$ and $k_z$. 
 
\begin{figure}[t!]
	\centering
		\includegraphics[width=\textwidth]{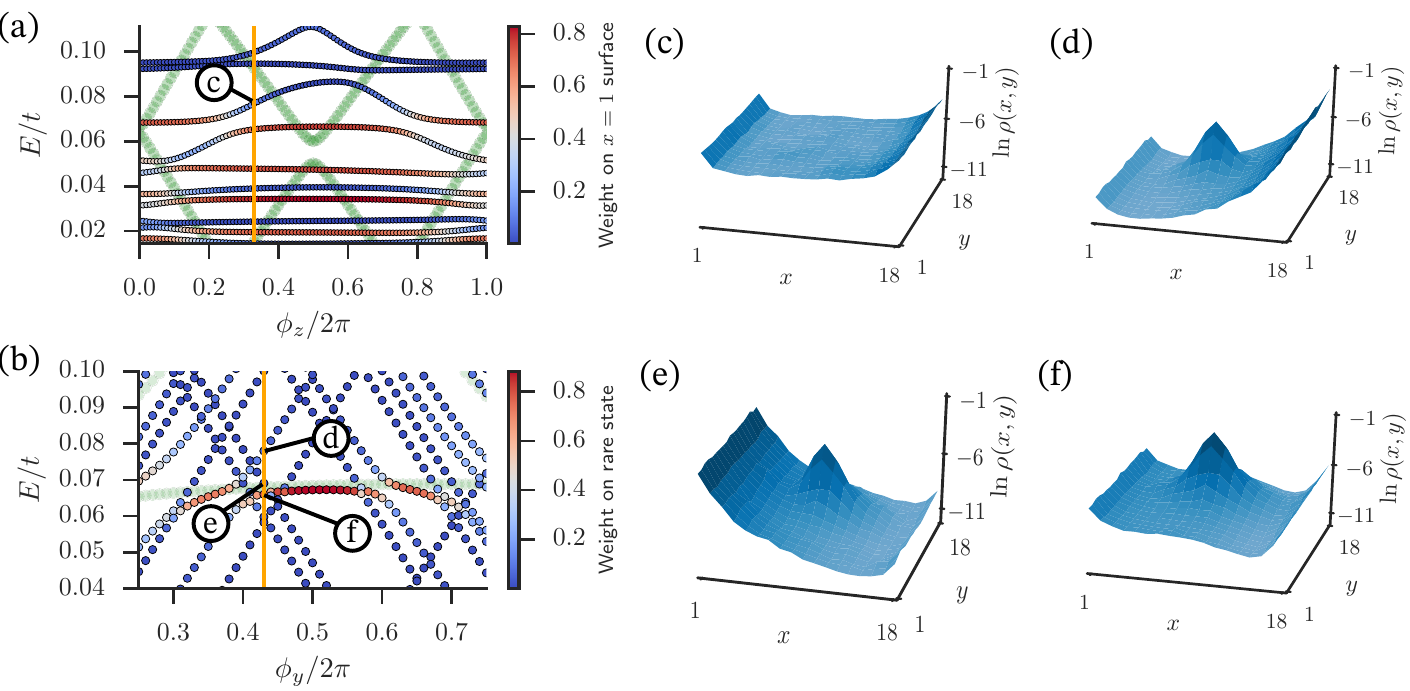}
	\caption{{\bf Fermi arc surface states hybridizing with perturbative and rare bulk states.} Energy eigenvalues as a function of the twist in the boundary condition (i.e. twist dispersions) versus a twist along $z$ for a perturbative state (a)  and a twist along $y$ for a rare state (b), in both cases periodic boundary conditions shown in green. In (a) we identify a perturbative hybridization between a bulk Weyl state and a topological Fermi arc surface state that is shown in (c), where as in (b) we see that the linearly dispersing Fermi arc surface state has hybridized with the weakly dispersing rare state. Corresponding hybridized wavefunctions are shown in (d), (e), (f) as labelled in (b) demonstrating that they are linear combinations of Fermi arc surface states  and quasilocalized rare states. Taken from Ref.~\cite{Wilson-2018}.}
	\label{fig:hybridize}
\end{figure}

 These surface states produce a metallic  density of states. Upon introducing disorder, these surface states broaden and develop a non-zero quasiparticle lifetime both perturbatively and non-perturbatively. Despite being quasi-two-dimensional they do not localize in the presence of weak disorder. Within perturbation theory these surface states remain bound to the surface in a power law fashion~\cite{gorbar_origin_2016}. Non-perturbative rare states in the bulk however hybridize with each state along the arc as they are broad in $k-$space due to the $1/r^2$ power law tail (as discussed in Sec.~\ref{sec:qp-excitations}). 
 As a result, the extensive bulk DOS fully hybridizes with the surface states in the large $L$ limit \cite{Wilson-2018}. The surface-bulk hybridization due to rare regions is shown in Fig.~\ref{fig:hybridize}.
 
\begin{figure}
	\centering
		\includegraphics[width=0.95\columnwidth]{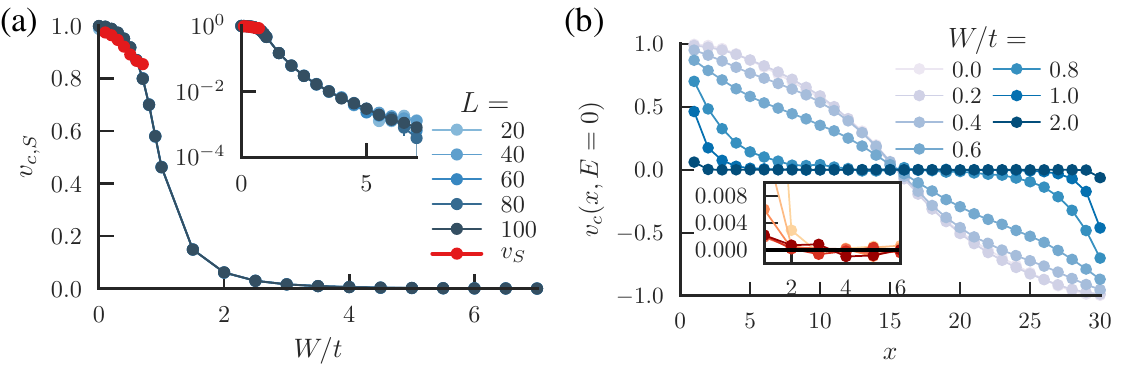}
	\caption{{\bf The chiral velocity of Fermi arc surface states in the presence of disorder.} 
	(a) The surface chiral velocity as a function of disorder strength $W$ for various system sizes as well as its value extracted from the pole of the surface Green function (in red). 
	(Inset) Surface chiral velocity on a log-scale showing that it remains non-zero even upon entry into the Anderson insulating phase, which it does by forming local current loops within the localization length. 
	(b) The average chiral velocity per sheet as a function of $x$ as in Eq.~\eqref{eqn:Jc}, showing that it becomes random in the bulk and remains on the edges to large disorder. Taken from Ref.~\cite{Wilson-2018}.}
	\label{fig:chiral_velocity}
\end{figure}

 Despite the surface-bulk hybridization the chiral velocity of the states remains remarkably robust. In the clean limit the chiral velocity is defined as ${\bf v}_c=\partial_{\bk_{\perp}}E_S(\bk_{\perp})$ and at zero energy is given by ${\bf v}_c=t \hat y$,  thus we can focus on the $y$-component in the following. In the presence of disorder, we generalize the definition of the chiral velocity in terms of derivative of a twist in the boundary condition ($\phi_y$), which is given by 
 \begin{equation}
     v_{c,S}=\mathrm{Tr}_{S(0)}(\partial H /\partial \phi_y|_{\phi_y=0})
 \end{equation}
 where $\mathrm{Tr}_{S(n)}$ is a trace over the sheet (in the $y-z$ plane) $n$ with $n=0$ being one of the two surfaces, and $-e\partial H /\partial \phi_y|_{\phi_y=0}=J_y$  is the  current operator along the $y$-direction. Generalizing this notion, we define a chiral velocity per sheet of the system $S(x)$ in a manner that can be computed with the KPM to yield 
 \begin{eqnarray}
v_{c}(x,E)= \frac{\mathrm{Tr}_{S(x)}(J_y\delta(E-H))}{\mathrm{Tr}_{S(x)}(-e \delta(E-H))},
\label{eqn:Jc}
\end{eqnarray}
for the Hamiltonian $H$.
Data for the surface chiral velocity as a function of disorder strength is shown in Fig.~\ref{fig:chiral_velocity}(a). 
Interestingly, while on a linear scale it appears that $v_{c,S}$ goes to zero at the Anderson localization transition in the bulk (that occurs near $W_l/t\approx 5.6-6.0$), when viewed on a log-scale (see inset) we see the chiral velocity remains non-zero but exponentially small as the model Anderson localizes. 
When resolving the chiral velocity per sheet, as see in Fig.~\ref{fig:chiral_velocity}(b), we see that the average remains concentrated to the surface as we increase the strength of disorder and at large disorder the chiral velocity in the bulk is random and hence averages out, whereas the chiral velocity on the surface remains non-zero. 
By analyzing the nature of the surface current it is possible to understand the non-zero chiral velocity in the Anderson insulating phase as a result of a non-zero current loops (into the bulk) that reside within a localization length of the surface.

 \subsection{Axial anomaly and chiral charge pumping}
 \label{sec:axial_anomaly}

 So far our analysis has focused on Weyl  semimetals in the presence of disorder.  We apply a constant magnetic field in the $z$-direction to the model in Eq.~\eqref{eqn:weylIS}, which we include
by Peierls substitution 
\begin{equation}
    t_y \mapsto t_y e^{-iBx}, \,\, t'_y \mapsto t'_y e^{-iBx},
\end{equation} 
for all sites. In order take periodic boundary conditions in the $x$ and $y$ directions we choose a gauge where 
\begin{equation}
    t_x \mapsto t_x e^{-iBL_x y}, \,\, t'_x \mapsto t'_x e^{-iBL_x y}
\end{equation} for the boundary hopping terms between the first and last sites (i.e. $x = 1$ and $x = L_x$, where $L_x$ is the system size in the $x$-direction). 
(Note that we are setting the lattice constant $a=1$, as well as $\hbar = e = 1$). Due to the periodic boundary conditions in $x$ and $y$ we can only access integer values of the magnetic flux quanta $\Phi_0 = h/e$.
\begin{figure}[t!]
	\centering
		\includegraphics[width=\textwidth]{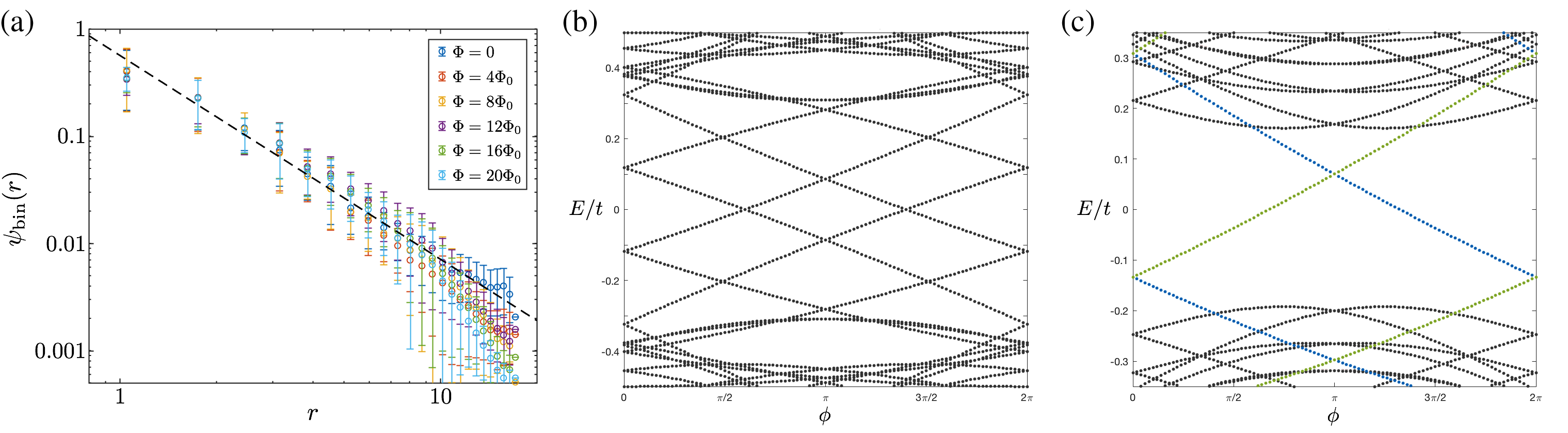}
	\caption{{\bf Magnetic field effects on rare state and band structure.} (a) A rare state in the presence of an orbital magnetic field of various strengths parameterized by the flux $\Phi$ for multiplies of the flux quanta $\Phi_0$ survives and does not affect the short range nature of the power law decay, though the longer range nature of the wavefunction is modified by the field. 
	At larger fields, when the magnetic length is of similar size as the rare region we expect this behavior to change, which is an interesting open question.   
	(b) For a twist along the $z$-direction (Eq.~\eqref{eqn:weylIS}) without a magnetic field, we clearly observe the two Weyl cones in the twist dispersion in the folded band structure. 
	(c) In the presence of a magnetic field two chiral Landau levels with opposite velocity develop (green and blue) along the direction of the magnetic field emanating out of the two Weyl points  with different helicities. Taken from Ref.~\cite{lee_chiral_2018}. }
	\label{fig:axial1}
\end{figure}

To demonstrate that rare eigenstates survive the addition of a magnetic field we first find a rare state with $B=0$ and then ``turn on'' the field as shown in Fig.~\ref{fig:axial1}(a). This demonstrates that the power-law bound rare eigenstates survive the addition of a magnetic field, which is quite natural as their local nature will only be affected by fields with a magnetic length of comparable size. We see in Fig.~\ref{fig:axial1}(a) that the rare state is modified at larger distances but the short distance power law tail remains robust for this range of magnetic fields.

The lowest Landau levels of the time-reversal broken Weyl model in Eq.~\eqref{eqn:weylIS} are chiral, emanating out of each Weyl node that disperse along the direction of the applied magnetic field (in this case the $z$-direction). To probe the chiral Landau levels in the absence of translational symmetry we apply a twist ($\phi_z$) along the $z$-direction and study how the low energy eigenstates near the chiral Landau levels disperse as a function of $\phi_z$. In the clean limit as shown in Fig.~\ref{fig:axial1} (c)  this produces two Landau levels of opposite chirality emanating out of the original Weyl points that can be seen in Fig.~\ref{fig:axial1} (b).

\begin{figure}[t!]
	\centering
		\includegraphics[width=0.95\columnwidth]{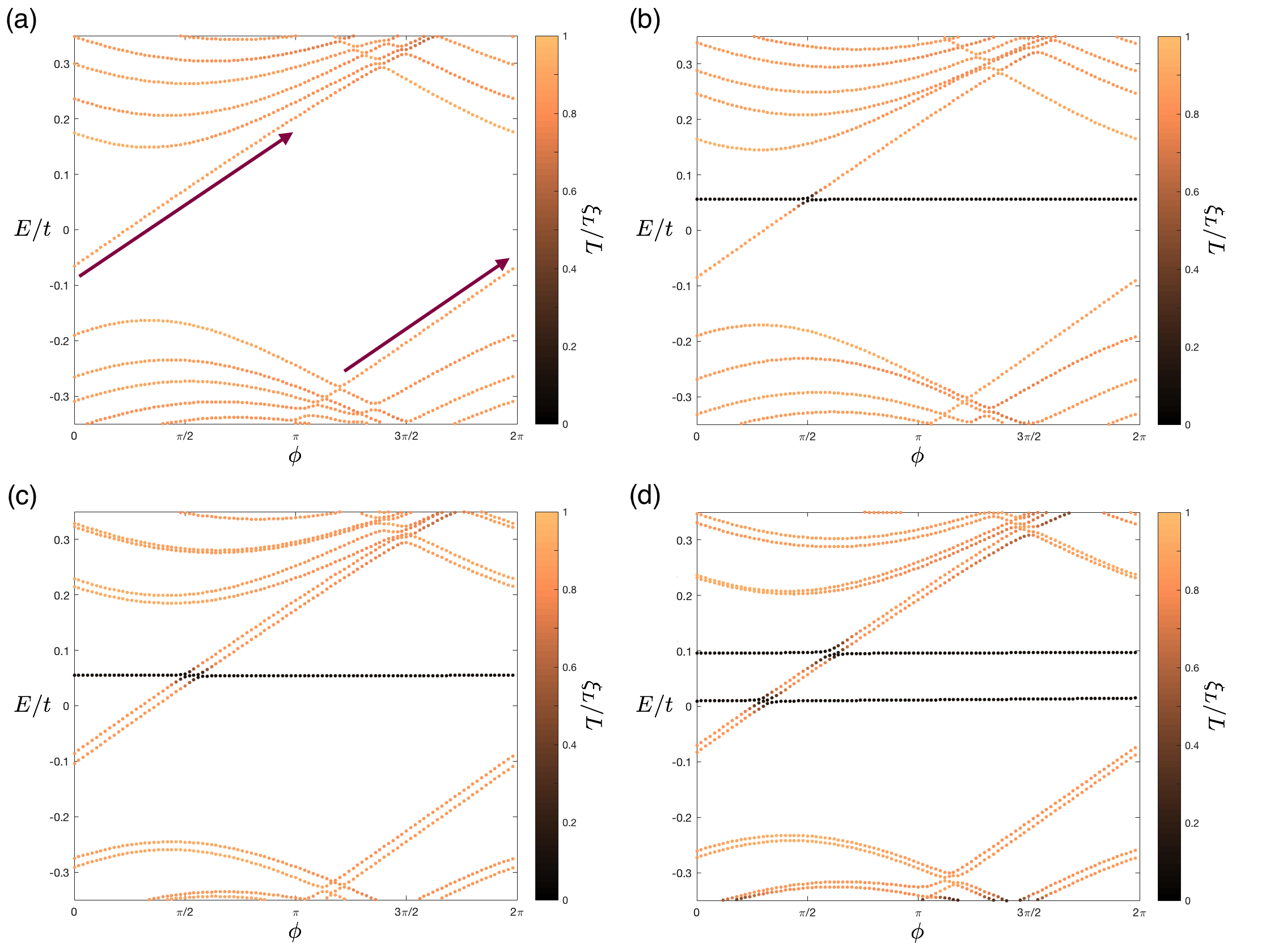}
	\caption{{\bf A single Weyl cone realized in the lattice model in Eq.~\ref{eqn:weylIS} and rare states in the presence of an orbital magnetic field $\Phi$.} The single Weyl node is only effective as we have added the non-local potential in Eq.~\eqref{eqn:nonlocal_potential} with $U_0=2t$ to ``push'' the other Weyl node to high energy. (a) The dispersion curve of the single chiral Landau level in the clean limit, the arrows denote the chiral band consistent with Eq.~\eqref{eqn:chiralband}. (b) A sample with a rare eigenstate and a magnetic field corresponding to one flux quanta $\Phi=\Phi_0$. The rare state  is weakly dispersing and it hybridizes with the chiral Landau level. Nonetheless, the topological charge pumping process following Eq.~\eqref{eqn:nonlocal_potential} remains satisfied. (c) The same rare sample with  two flux quanta $\Phi=2\Phi_0$, now the rare state only disrupts the chiral nature of the band for one of the chiral Landau levels. (d) A sample with two rare regions that produces a bi-quasi-localized wavefunction that is able to hybridize with both of the chiral Landau levels but does not destroy the topological nature of the bands. The color indicates $\xi_L=(\sum_{\mathbf r} |\psi_\mathbf{r}(E)|^4)^{-1/3}$ which is a localization length; when $\xi_L\sim L$ the state is maximally delocalized. Taken from Ref.~\cite{lee_chiral_2018}.  }
	\label{fig:axial3}
\end{figure}

We can further deform the model to access the physics of a single cone by applying a non-local potential. In particular, we add a momentum dependent potential $U({\bf k})$ to the model in Eq.~\eqref{eqn:weylIS} (in the parameter space of two Weyl nodes), where 
\begin{equation}
    U({\bf k})  = \begin{cases} 0, &  \mathrm{for}\;\; 0 \leq k_z < \pi,
    \\
     U_0, & \mathrm{for}\;\; \pi \leq k_z < 2\pi.
    \end{cases}
    \label{eqn:nonlocal_potential}
\end{equation} 
and by focusing on low energies we can ignore the other cone that has been pushed to much higher energies that we take to be $U_0=2t$. As a result, in the presence of a magnetic field at low energy we are now effectively probing a single Weyl point as shown in Fig.~\ref{fig:axial3}(a).

In the following we will distinguish between chiral bands, where the topological charge pumping process can occur, from non-chiral bands where this has been destroyed by examining the velocity of the each eigenvalue $E_n$, $v(\phi)=dE_n/d\phi$ (where $\phi$ is the twist along $z$, dropping the subscript in this section)  through one pumping cycle. Namely, the bands are chiral if 
\begin{equation}
    \delta E = \int_0^{2\pi} v(\phi)d\phi =E(2\pi)-E(0) \neq 0,
    \label{eqn:chiralband}
\end{equation} 
whereas trivial bands will always have $\delta E=0$.

Focusing on the  limit of a single Weyl node and a field strength corresponding to one flux quanta, as shown in Fig.~\ref{fig:axial3}(a), we see the non-local potential has successfully pushed the other Weyl point to higher energy. As found in Fig.~\ref{fig:axial3}(b) the non-dispersing band due to a single rare region hybridizes with the chiral Landau level. However, it does so in manner that retains the chiral nature of the band  as defined in Eq.~\eqref{eqn:chiralband} by wrapping around the mini-BZ to continue the charge pumping process and the axial anomaly survives. In other words, the presence of a rare state for a single Weyl node only slows down the charge pumping process but does not destroy it.
It is interesting to then thread two flux quanta through the same rare sample as shown in Fig.~\ref{fig:axial3}(b), where the rare state produces one staircase while the other chiral Landau level persists almost unaffected (apart from near their hybridization). In Fig.~\ref{fig:axial3}(d) we consider two flux quanta and a sample with two rare regions [similar to Fig.~\ref{fig:WFs}(c)] and now each rare state hybridizes with the two Landau levels while still allowing the charge pumping process to survive.

\begin{figure}[t!]
	\centering
		\includegraphics[width=\textwidth]{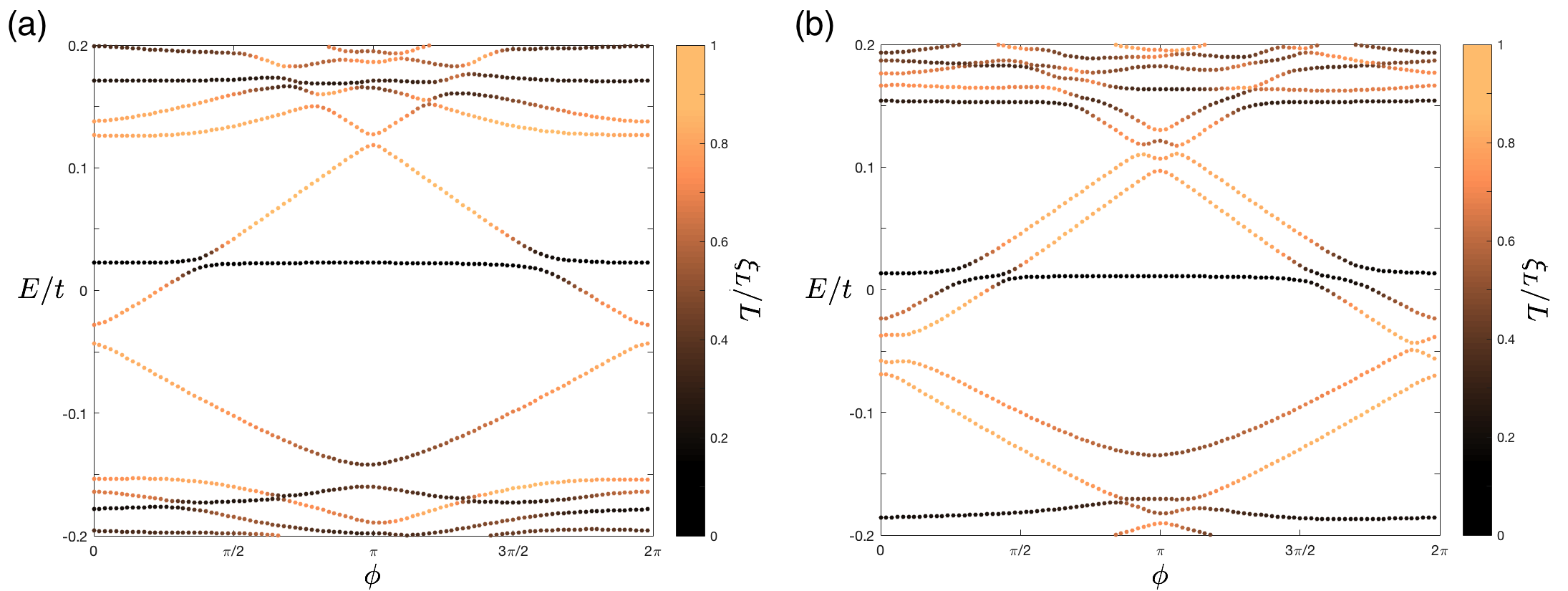}
	\caption{{\bf A rare state in the presence of an orbital magnetic field for the model in Eq.~\eqref{eqn:weylIS} with two Weyl nodes}.(a) For one flux quanta $\Phi=\Phi_0$ a rare state hybridizes the two chiral Landau levels at low energy near the Weyl node destroying the topological charge pumping process. (b) For two flux quanta $\Phi=2\Phi_0$ and one rare state one renormalized chiral Landau level remains.  Taken from Ref.~\cite{lee_chiral_2018}. }
	\label{fig:axial4}
\end{figure}

However, if we remove the non-local potential so that we are considering a physically relevant model with two Weyl points at the same energy then the rare state hybridizes the two distinct chiral Landau levels and destroys the topological charge pumping, as shown in Fig~\ref{fig:axial4}(a). While this also occurs perturbatively due to internode scattering, the rare state induces this effect at much lower energy at weak disorder (e.g., notice where the two chiral Landau levels level repel each other at $\phi=\pi$ at much higher energy then the rare state is occurring at). 
On the other hand, as we increase the flux quanta, other chiral Landau levels will be induced and therefore a single rare state is not sufficient to hybridize all four chiral Landau levels. Instead one pair survives that allows for the chiral charge pumping to persist for this finite size sample, Fig~\ref{fig:axial4} (b). However, in the thermodynamic limit the number of flux quanta scale as $\sim L^2$ and the density of rare states scales as $\sim L^3$ and thus rare regions will inevitably destroy the charge pumping process for any number flux quanta.
 
 In the limit of larger disorder strength so that the model is sufficiently deep in the diffusive metal, phase a non-linear sigma model analysis is appropriate \cite{altland_effective_2015,altland_theory_2016}. 
 In this limit there is no longer any sense of ``band'' to talk about, yet the chiral anomaly and the charge pumping process survives in the limit of a single Weyl node.
 While this study has discussed the microscopic effect of rare regions on the chiral charge pumping process it has not resolved the nature of this will appear in the transport and the magneto-resistance. It will be interesting to explore these questions in future  work.
 
\section{Particle-hole symmetry}
\label{sec:ph_symmetry}

\begin{figure}[t!]
	\centering
		\includegraphics[width=\textwidth]{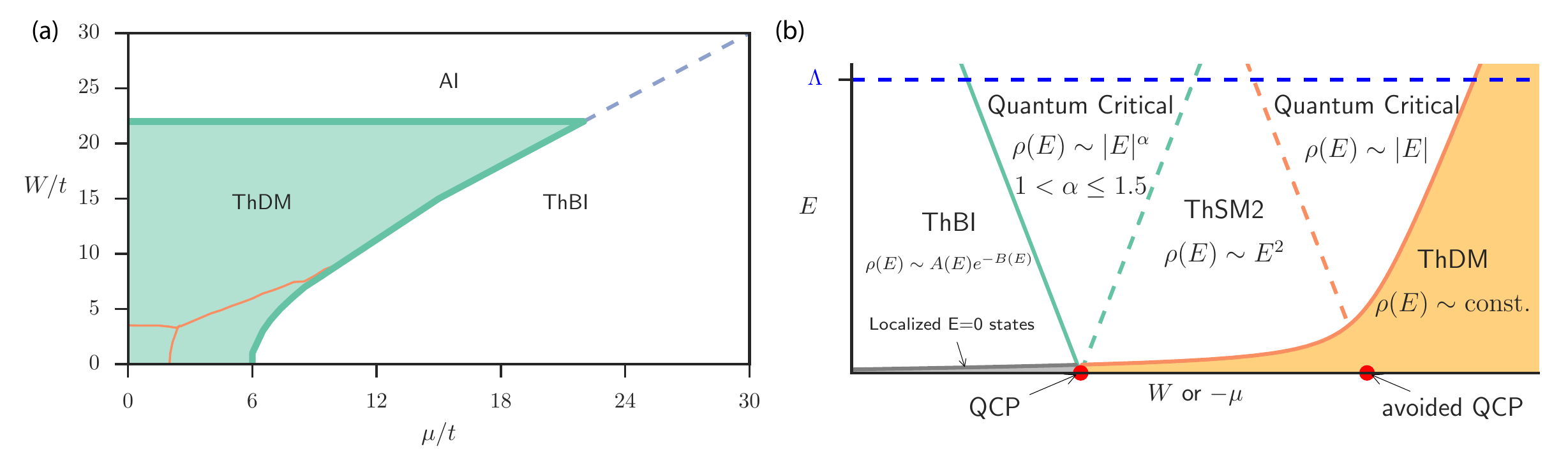} 
	\caption{{\bf Phase diagram of a disordered three-dimensional $p+ip$ superconductor in Eq.~\eqref{eq:pipHamiltonian}.} (a) As rare regions destabilize the vanishing density of states of the $p+ip$ superconductor the phase diagram at the center of the band ($E=0$) has only two phases at finite disorder that are a diffusive metal and Anderson insulating phase (that is smoothly connected to the band insulator). The orange lines denote the cross over boundaries between 2 and 4 Weyl nodes (varying $\mu$) and the AQCP (as a function of $W$).
	(b) Phase diagram in terms of the average density of states across the band (or Anderson) insulator to diffusive metal phase transition and the cross over at the avoided transition. Note that the true Anderson localization transition shown here with a QCP label is not critical in the average density of states as it is made up of localized Lifshitz states~\cite{yaida_instanton_2016}. Figures taken from Ref.~\cite{Wilson-2017}.}
	\label{fig:phasediagramSC}
\end{figure}

We now turn to two models that have an exact particle-hole symmetry. 
The first of which is the superconducting analog of Eq.~\eqref{eqn:weylIS}, and thus potential disorder enters as a particle-hole symmetric term realizing a model in class D  that was studied in detail in Ref.~\cite{Wilson-2017}. The second model we consider is a bipartite random hopping model of the form that appears in Eq.~\eqref{eqn:weylTR} that falls into class BDI. The results obtained for the random hopping model are new, being reported here for the first time.
These problems represent two distinct examples of disordered Weyl semimetals in chiral symmetry classes of purely off-diagonal matrices where the zero energy state in the center of the band plays a special role. In the models we consider, the Weyl node energy sits precisely at the symmetric band center to see if particle-hole symmetry can somehow  ``protect'' the Weyl state from disorder. However, as we will see below rare regions still dominate the behaviour of the Weyl node and the avoidance is only enhanced by the chiral symmetry.
 
 \subsection{A $p+ip$ Weyl superconductor}

 In Ref.~\cite{Wilson-2017}, a three-dimensional model for a $p+ip$ superconductor was investigated in the presence of disorder. Here, we work under the assumption of a non-zero pairing amplitude that we do not determine self-consistently. The model Hamiltonian is given by
 \begin{equation} 
H_{\mathrm{SC}} = \sum_{\mathbf r,\hat{\nu}} \;  \left[t \; c_{\mathbf r+\hat{\nu}}^{\dagger} c_{\mathbf r} + i \Delta_{\hat{\nu}} \; c_{\mathbf r+\hat{\nu}}^{\dagger} c_{\mathbf r}^\dagger + \mathrm{h.c.}  \right] 
+  \sum_{\mathbf r} (V(\mathbf r) -\mu) c_{\mathbf r}^{\dagger} c_{\mathbf r},  
  \label{eq:pipHamiltonian}
\end{equation}
where $\hat{\nu}=\pm \hat x,\pm \hat y, \pm \hat z$, the $p_x+ip_y$ superconducting gap is given by $\Delta_{\hat{x}}= \Delta, \; \Delta_{\hat{y}}=-i \Delta, \; \Delta_{\hat{z}}=0$, and $2\Delta$ is the maximum size of the superconducting gap for the clean model. In the absence of disorder we can rewrite Eq.~\eqref{eq:pipHamiltonian} using Nambu spinors $\Psi_{\br}=(c_{\br} \,\,\, c_{\br}^{\dag})^T$ and the clean part of model is then given by Eq.~\eqref{eqn:weylIS} (with $\psi_{\br}$ replaced with $\Psi_{\br}$) and this results in the particle hole symmetric disorder
\begin{equation}
    H_{\mathrm{disorder}}=\sum_{{\bf r}} \Psi_{{\bf r}}^{\dag}V({\bf r})\tau_z\Psi_{{\bf r}}
\end{equation}
 where $\tau_z$ is the $z$-Pauli matrix in Nambu space.
 
 \begin{figure}[t!]
	\centering
		\includegraphics[width=\textwidth]{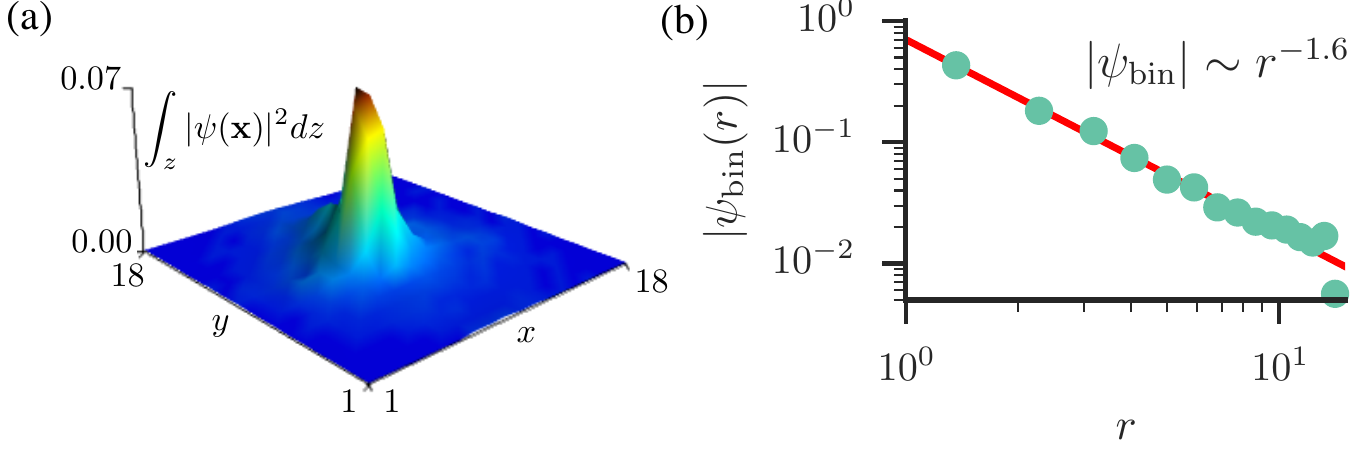} 
	\caption{
	{\bf Rare eigenstate in the $p+ip$ superconductor in Eq.~\eqref{eq:pipHamiltonian}.} The projected probability density of the wavefunction in the $x-y$ plane displaying a quasi-localized rare state. 
	(b) The rare state displays a clear power law decay from its maximal value with an exponent in reasonable agreement with the analytic expectation. Despite the presence of particle-hole symmetry the nature of the rare states look qualitatively similar to the case of potential disorder. Taken from Ref.~\cite{Wilson-2017}.}
	\label{fig:rareWFSC}
\end{figure}

The phase diagram of this model is shown in Fig.~\ref{fig:phasediagramSC}(a) in the space of $\mu-W$. 
This model allows for a much larger multitude of clean band structures to consider (such as anisotropic nodal points at $|\mu|=2t$). 
In addition to nodal superconductors in the clean limit and a diffusive thermal metal of the BDG excitations at finite disorder, this model also hosts a thermal band insulator of BDG quasiparticles at large $\mu$, which in the presence of disorder is smoothly connected to a thermal Anderson insulating phase that sets in at large disorder strength. The orange vertical line marks the anisotropic semimetal point that separates the 4 and 2 Weyl node regimes. The horizontal and diagonal orange line marks the avoided transition that eventually merges with the band insulator transition. The transition between the diffusive thermal metal and thermal Anderson insulator is a true Anderson localization transition of the BDG quasiparticles that will appear in level statistics and wavefunction properties but is not observable in the density of states. While perturbatively a disorder driven transition is also predicted along the boundary between the diffusive thermal metal and thermal band insulator, this is also rounded out into a crossover. This enriches the phenomena of the avoided transition into Fig.~\ref{fig:phasediagramSC}(b).

\begin{figure}[t!]
	\centering
		\includegraphics[width=\textwidth]{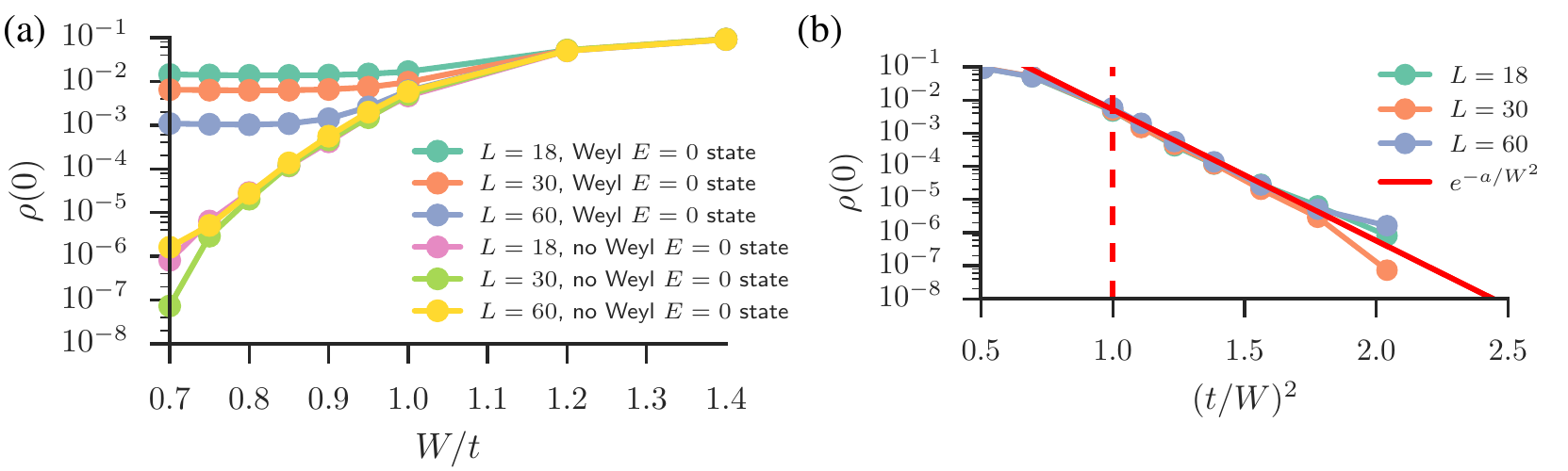} 
	\caption{
	{\bf Zero energy DOS for the $p+ip$ model in Eq.~\eqref{eq:pipHamiltonian}.}
	Utilizing twisted boundary conditions allows us to compute the rare region contribution to the density of states which is converged in system size and orders of magnitude below the case of periodic boundary conditions with $L$ that produces a large finite size effect due to the Weyl peak at zero energy. (b) The converged zero energy density of states plotted on a log scale versus  $(t/W)^2$ displaying that the rare region scaling form works over several orders of magnitude of the DOS. Taken from Ref.~\cite{Wilson-2017}.
	}
	\label{fig:DOS-SC}
\end{figure}

Focusing on the limit of four Weyl cones (see Fig.~\ref{fig:phasediagramSC}), the model possesses an AQCP due to rare regions. In Fig.~\ref{fig:rareWFSC}, we show a rare state that is power law bound with an exponent close to the analytic expectation in Eq.~\eqref{eqn:rareDOS}. In Fig.~\ref{fig:DOS-SC} we show the computed rare region contribution to the DOS being well fit by the rare region form in Eq.~\eqref{eqn:rareDOS} over several overs of magnitude.  
The density of states remains an analytic function and the peak in $\rho''(0)$ is saturated.

A major distinction between the current limit and potential disorder is once $\rho(0)$ is of order $O(t)$, an antilocalization peak appears in the low energy DOS. This can be captured within a non-linear sigma model analysis once the stiffness (i.e.\ $\rho(0)$) is of order one; it yields an energy dependent DOS 
that goes like 
\begin{equation}
    \rho(0)-\rho(E)\sim \sqrt{|E|}
    \label{eqn:antilocal}
\end{equation}
 near $E=0$ \cite{senthil_quasiparticle_2000}.
 
 \subsection{Beyond Linear Touching Points}
 \label{sec:beyond_linear_touching}
 
\begin{figure}
	\centering
		\includegraphics[width=0.6\columnwidth]
		{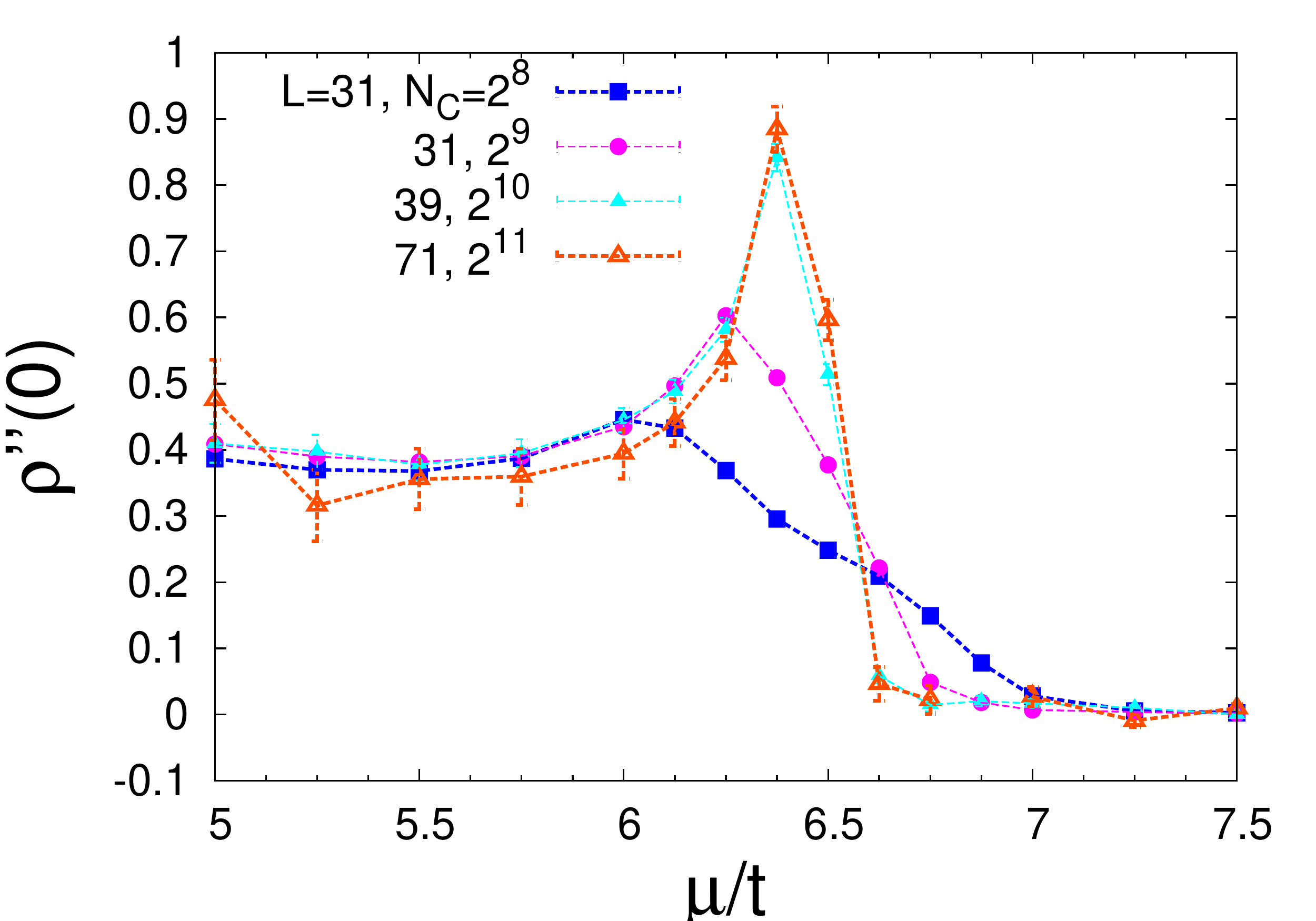}
	\caption{{\bf Rounding out the non-analytic behavior in the DOS across the diffusive metal-to-band insulator transition in the $p+ip$ superconductor.} The anisotropic nodal dispersion produces a non-analytic DOS $\rho(E)\sim |E|^{3/2}$ with a divergent $\rho''(0)\sim |\mu-\mu_I|^{-1/2}$ that is rounded out by rare regions that produce a smooth and analytic average DOS; as one can see since $\rho''(0)$ is saturated in $L$ and $N_c$. Taken from Ref.~\cite{Wilson-2017}.}
	\label{fig:rhopp_bandedge}
\end{figure}

 It is a natural question to ask: What the impact of rare regions is on other nodal band structures beyond just Dirac points that also maintain the perturbative irrelevance of disorder? There are now a large number of examples of topological semimetals \cite{xu_structured_2015,bradlyn_beyond_2016,wieder_double_2016} and it is an exciting question to study the effects of rare regions on their low energy excitations. 
 One clear example of this is  the anisotropic nodal band structure at the transition between a Weyl semimetal and a band insulator, i.e.\ by traversing the phase diagram in Fig.~\ref{fig:phasediagramSC} (a) in $\mu$ at $W=0$. 
 At this location the low energy density of states goes like $\rho(E)\sim |E|^{3/2}$ and is non-analytic (see Sec.~\ref{sec:TRbrokenWeyl}) and disorder is perturbatively irrelevant, and the critical theory is sketched in Fig.~\ref{fig:phasediagramSC}(b). 
 In particular, as we tune $\mu$ in the clean limit, we have $\rho''(0)\sim |\mu-\mu_I|^{-1/2}$. 
 As shown in Fig.~\ref{fig:rhopp_bandedge}, this non-analytic behavior in the DOS is rounded out by rare regions and the transition between the Weyl semimetal and the band insulator is no longer sharp in the average DOS.
 
On the other hand, the transition between the Weyl semimetal and the band insulator is indeed a sharp transition with regards to the nature of the wave-function. 
States in the band insulator are exponentially localized Lifshitz states that round out the sharpness of the spectral band gap. 
However, transport and wave function probes (such as the inverse participation ratio or the typical density of states) are sensitive to this realizing a true quantum phase separating a diffusive metal and an Anderson insulator. 
Due to the anisotropic nodal band structure and the irrelevance of disorder, an avoided quantum phase transition also appears here in the average DOS, see Figs.~\ref{fig:phasediagramSC} and \ref{fig:rhopp_bandedge}.

\subsection{Random hopping model}
  
A simpler model with particle-hole symmetry can be realized with a random hopping model with only a single term in the model. This will allow us to interpolate between weak and strong randomness as well directly study the interplay of rare regions and the ``pile up'' of low energy states due to  antilocalization effects.
  
  In this section, we present new results on the model in Eq.~\eqref{eqn:weylTR} with a random hopping, namely
\begin{equation}
H_{\mathrm{Weyl}}^{RH}= \sum_{{\bf r},{\mu}={x},{y},{z}}\left( i t_{\mu}(\br)\psi_{{\bf r}}^{\dag}\sigma_{\mu} \psi_{{\bf r}+\hat{\mu}} + \mathrm{H.c.}\right),
\label{eqn:RH}
\end{equation}
where $t_{\mu}(\br)$ is taken to be a random variable along each bond (labeled by $\br$ and $\mu$). We parametrize $t_{\mu}(\br)$  with mean $\sqrt{1-W^2}$ and standard deviation $W$ so that the limit of purely random hopping is accessible at $W=1$. Using the symmetries of the hopping model derived in Sec.~\ref{sec:inversionWeyl} [see Eq.\eqref{eq:symmetry}] the current bipartite random hopping model falls into the chiral orthogonal BDI Altland-Zirnbauer class \cite{altland_nonstandard_1997}.
 
\begin{figure}[t!]
	\centering
\includegraphics[width=\textwidth]{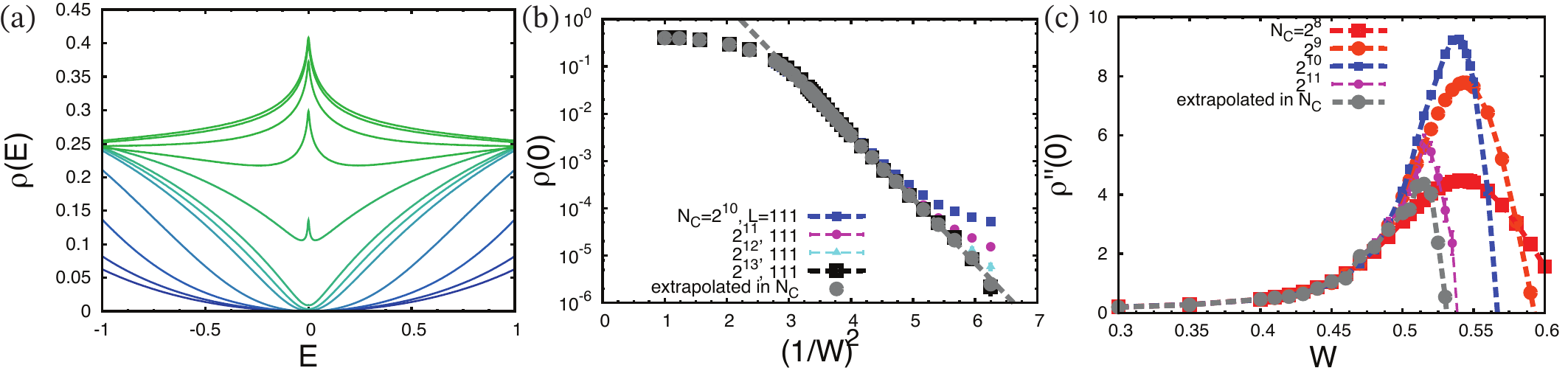}
	\caption{
	{\bf Results demonstrating a strong avoidance in the random hopping model in Eq.~\eqref{eqn:RH}.} Here we focus on a large system size of $L=111$ using the KPM to eliminate any finite size effects at these KPM expansion orders. 
	(a) The density of states $\rho(E)$ as a function of energy $E$ for various disorder strengths starting from $W=0.1$ (blue) to $W=1$ (green) computed for a KPM  expansion order $N_C=2^{13}$ averaged over 1000 samples each with a random twist in the boundary condition. 
	We see a clear anti-localization peak [with the form given in Eq.~\eqref{eqn:antilocal}] appear at low energy roughly when the density of states appears to lift off from zero on this linear scale. 
	(b) The zero energy density of states versus $(1/W)^2$ for various expansion orders as well as the value extrapolated in $N_C$ showing a clear rare region scaling consistent with Eq.~\eqref{eqn:rareDOS} over almost five orders of magnitude (fit shown as a grey dashed line). 
	(c) The second derivative of the density of states $\rho''(0)$ versus disorder strength $W$ to probe the analytic properties of the density of states. 
	The antilocalization peak is a separate nonanalyticity that is generated when the zero-energy density of states is sufficiently large. 
	Here, we see that leads to an additional rounding of the peak in $\rho''(0)$ signalling an analytic density of states at the avoided transition. For larger disorder strengths the antilocalization peak takes over, causing a massive, negative $\rho''(0)$.
	}
	\label{fig:RH-model}
\end{figure}

 The average DOS as a function of energy of the random hopping model is shown in Fig.~\ref{fig:RH-model}(a) for several values of $W$. The appearance of the antilocalization peak (described by Eq.~\eqref{eqn:antilocal} \cite{gade_anderson_1993}) clearly appears right after the AQCP (which occurs near $W_c\approx 0.525$) and continues to sharpen. Examining the zero energy DOS at the Weyl node energy, is shown in Fig.~\ref{fig:RH-model}(b) demonstrating an excellent fit to the non-perturbative rare region form in Eq.~\eqref{eqn:rareDOS}. In order to analyze our results to obtain an $N_C$-independent estimate of $\rho(0)$ and $\rho''(0)$ we utilize the recent analysis we put forth in Ref.~\cite{wilson_avoided_2020-1} that utilizes the convolution of the Jackson kernel with the analytic scaling form of the density of states in Eq.~\eqref{eqn:dos-analytic} to obtain 
 \begin{equation}
     \rho_{N_C}(0)=\rho(0)+\frac{1}{2}\rho''(0)\left(\frac{\pi \Delta}{N_C}\right)^2+\cdots
     \label{eqn:extrap}
 \end{equation}
 where $\rho_{N_C}(0)$ is the KPM data and $\Delta$ is the bandwidth of the model. The result of the extrapolation is shown in Fig.~\ref{fig:RH-model}(b) and (c) as grey circles demonstrating the density of states is converged across this range of $W$.
 
Coming to the nature of the avoidance we show results for $\rho''(0)$ in Fig.~\ref{fig:RH-model}(c) for various KPM expansion orders that are converged at this system size ($L=111$).
Distinct from our previous discussions, due to antilocalization effects $\rho''(0)$ changes sign and becomes negative for $W \gtrsim 0.55$. 
The peak at the AQCP is strongly rounded out; while it initially sharpens for increasing $N_C$, it  eventually decreases due to antilocalization effects. In order to try and extract an $N_C$-independent estimate we extrapolate the zero energy density of states across the range of $N_C=2^9-2^{13}$ using Eq.~\eqref{eqn:extrap} to obtain the grey circles indicating a very strong rounding in the large $N_C$ limit. In summary, our results are consistent with random hopping also destabilizing the Weyl semimetal phase and an interesting interplay between the perturbative transition, rare regions, and antilocalization effects lead to an even stronger rounding of $\rho''(0).$

 \section{A single Weyl node in the continuum}
 \label{sec:single_Weyl_node}
 
The numerical models studied thus far included band curvature and an even number of Weyl nodes in the Brillouin zone.
It could reasonably be argued that scattering between opposite chirality nodes and/or band curvature effects could produce rare resonances that we see.
Additionally, work by Buchhold et al.~\cite{buchhold_vanishing_2018,buchhold_nodal_2018} has suggested that $\rho(0)=0$ identically due to individual rare resonances and in any reasonable disorder scheme.
As a consequence, criticality could be restored.
It is thus important to study the perfectly linear, single Weyl node to determine if avoided criticality persists, and in this section we show explicitly that it does, suggesting rare resonances as described in Sec.~\ref{sec:rare_resonances} play a significant role in the physics as we have described.
 
The single Weyl node Hamiltonian in the continuum takes the form
\begin{equation}
    H=-iv\bm{\sigma}\cdot\nabla + V(\br)
    \label{eqn:singlecone}
\end{equation}
where the potential is Gaussian distributed with zero mean and 
\begin{equation}
    \langle V(\br+\bR)V(\br) \rangle = W^2 e^{-R^2/\xi^2}.
\end{equation}
Without loss of generality we set $v=1$ and $\xi=1$ in the following. 
To simulate this model, a discretization of momentum space is required. 
However, the discretized grid or ``lattice'' is arbitrary and our results should not depend on it. 
Therefore, we consider two different discretizations of the continuum in momentum space the simplest being a simple cubic lattice and the second is a face centered cubic (FCC) lattice (which provides the most dense sphere packing in 3D). 
As demonstrated in Refs.~\cite{sbierski_strong_2019,wilson_avoided_2020-1} it is possible to still implement sparse matrix-vector techniques by working with vectors in momentum space and whenever we act with the potential (that is defined in real space) we use a series of fast Fourier transforms (and their inverse). 
For a matrix of size $\mathcal{N}$ this increases the computation cost from $\mathcal{N}$ to $\mathcal{N} (\log \mathcal{N})^3$ and the matrix-vector approaches remain fast.
 
\begin{figure}
    \centering
    \includegraphics[width=\textwidth]{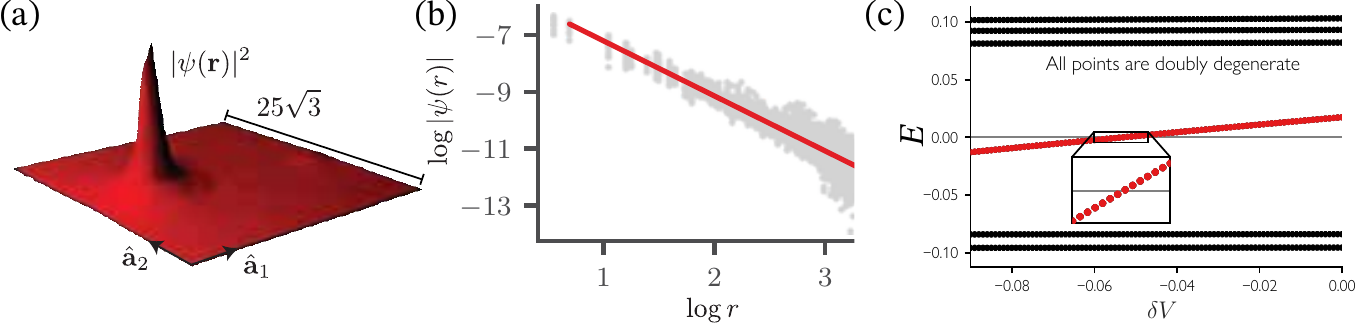}
    \caption{A rare state found in the fcc model in Eq.~\eqref{eqn:singlecone} with system size $L=25\sqrt{3}$ and cutoff $\Lambda=\pi/\sqrt{2}$ and disorder strength below the avoided quantum critical point of $W=0.7$ at energy $E=0.0168$. 
    (a) The probability density of the rare wave function for a cut through the real-space bcc lattice where $\mathbf r= n_1 \mathbf a_1 + n_2 \mathbf a_2 + 22 \mathbf a_3$ $\mathbf a_1=(1,-1,1)$, $\mathbf a_2=(-1,1,1)$, and $\mathbf a_3 = (1,1,-1)$.
    (b) A scatter plot of the probability density (gray) and the best fit line (red) with power-law decay $|\psi(\mathbf r)| \sim 1/r^{1.94}$. 
    (c) Adding a spherical potential to the disorder potential at the maximum of $|\psi(\mathbf r)|$ with radius two and strength $\delta V$, we can tune the rare state energy through zero energy. Taken from Ref.~\cite{wilson_avoided_2020-1}.}
    \label{fig:scw_rarestate}
\end{figure}

While there are prior numerical results in the limit of a single Weyl cone, no rare states were identified. 
An example of a rare state found using the FCC discretization is shown in Fig.~\ref{fig:scw_rarestate}(a). 
This displays the expected power law going like $\psi_{\mathrm{rare}}(\br)\sim 1/|\br- \br_0|^{1.94}$ (Fig.~\ref{fig:scw_rarestate}(b)) in excellent agreement with Eq.~\eqref{eqn:rarestate}.
The energy of this state is $E=0.0168$ and close to zero, being within the smallest finite size gap in the middle of the band.
We can tune the energy of this rare state to precisely zero by adding a small, local spherical potential centered about the rare region without affecting the rest of the spectrum as shown in Fig.~\ref{fig:scw_rarestate}(c).

\begin{figure}
    \centering
    \includegraphics{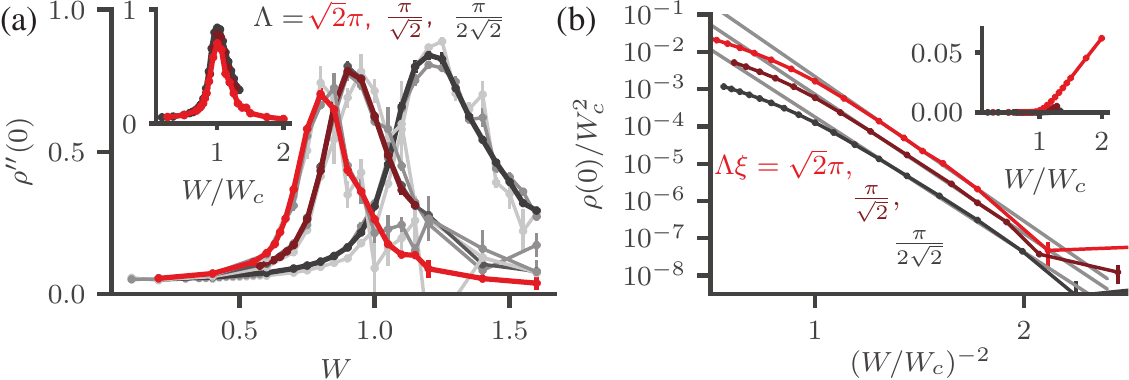}
    \caption{{\bf The avoidance survives a single, perfectly linear Weyl cone.}
    Fitting $\rho(0)$ as a polynomial in $1/N_C$ allows us to extract $\rho(0)$ and $\rho''(0)$ as explained in Ref.~\cite{wilson_avoided_2020-1}; the results at different sizes and momentum space cutoffs are presented. Multiple sizes are shown at a given cutoff: light-to-dark gray curves represent different sizes $L\Lambda\sqrt{2}/\pi = 32\sqrt{3},64\sqrt{3},128\sqrt{3},128\sqrt{3}$ (and $L=160\sqrt{3}$ when $\Lambda=\pi/\sqrt{2}$). 
    (a) $\rho''(0)$ is drifting as a function of the cutoff, but the size of the avoidance remains the same. In fact, the drift due to cutoff can be understood by a simple generalization of the self-consistent Born approximation \cite{wilson_avoided_2020-1}. (Inset) shows the data normalized by the peak of $\rho''(0)$, where the avoided critical point $W_c$ occurs.
    (b) $\rho(0)$ fits the rare region form well. Other system size data is omitted since it lies directly on top of the shown curve. The gray curves represents a fit to the rare region equation \eqref{eqn:rareDOS}. Inset is the same data on a linear scale. Taken from Ref.~\cite{wilson_avoided_2020-1}.}
    \label{fig:scw_avoidance}
\end{figure}

Using this model, we have removed all of the effects of band curvature and multiple Weyl nodes present in the simulation. 
As a result, we can now characterize the strength of avoidance in the limit of a perfectly linear single Weyl cone. 
As shown in Fig.~\ref{fig:scw_avoidance}, the avoidance remains strong with Gaussian disorder independent of the cut-off and discretization used. 
This is indicated by an $N_C\rightarrow\infty$ extrapolated $\rho''(0)$ which further appears saturated with system size in Fig.~\ref{fig:scw_avoidance}(a) and a similarly extrapolated in $N_C$ and converged in $L$ $\rho(0)$ in accordance with Eq.~\eqref{eqn:rareDOS}.
Thus, our results demonstrate the AQCP survives in the limit of a single Weyl cone. 
This also directly suggests that a single Weyl cone is unstable to infinitesimal disorder due to the appearance of a finite density of states.
 
 \section{Discussion}
 
 In this review we have examined numerical results on rare region effects in Dirac and Weyl semimetals. While it is possible to tune the probability of generating rare events we discussed how it is not possible to exactly remove rare events in the presence of randomness. We have shown a great deal of numerical evidence in the presence and absence of particle hole symmetry, as well as in the BdG spectrum that disorder generically destroys the semimetal phase in three-dimensions and the putative transition is avoided, which is rounded out into a crossover. It will be  interesting to see how these effects appear in transport. In addition to the non-zero conductivity expected at the Weyl node the nature of the anomalous Hall effect in the presence of rare regions in disordered time reversal broken Weyl semimetals \cite{Shapourian-2015,kobayashi_ballistic_2020} and the photogalvanic effect \cite{konig_photogalvanic_2017} in inversion broken Weyl semimetals are both worthy of further investigation.
 
 The effects of rare eigenstates on the topological properties of Weyl semimetals have also been discussed. The destabilization of Fermi arc surface states due to hybridizing with non-perturbative rare eigenstates was demonstrated in conjunction with a remarkably robust surface chiral velocity. In the presence of magnetic fields rare eigenstates survive and destroy the  chiral charge pumping process that arises due to the axial anomaly in the presence of parallel electric and magnetic fields. Interestingly, in the limit of a single Weyl cone this charge pumping process survives non-perturbative rare region effects. In future work, it is important to connect these observations with dynamical transport properties in the presence of electro-magnetic fields~\cite{burkov_dynamical_2018,cheng_probing_2019} to understand the role of effects of disorder on the chiral lifetime.

Last, we investigated the limit of a single Weyl cone with a linear dispersion numerically. This study was motivated by recent field theoretic results that have argued the Weyl semimetal phase is stable to rare region effects. However, the exact numerical simulations presented in Sec.~\ref{sec:single_Weyl_node} is at odds with this prediction as they are consistent with an avoided quantum criticality that implies the single Weyl node is unstable to disorder. These discrepancies between the field theory expectations and the numerics should be of central interest in future studies (Ref.~\cite{pires_randomness_2020} offers a one resolution, for instance).
 
  A wide open direction of research is developing a theoretical understanding of the role of interactions in the presence of rare region effects. Not only does this pertain to strongly correlated Weyl semimetals but also the formation of the Kondo-Weyl semimetal ground state may be affected. It will be exciting to determine the interplay of these effects in future work.

 We close with a brief discussion of a connection between this avoided criticality three-dimensions \cite{Pixley-2018} and the magic-angle effect in the band-structure of twisted bilayer graphene in two-dimensions \cite{fu_magic-angle_2020}. 
 By replacing the random potential by a three-dimensional, quasiperiodic potential, rare events are removed. 
 As a result, a true quantum phase transition between a Weyl semimetal and a diffusive metal can be stabilized \cite{mastropietro_stability_2020}.
 As the Weyl semimetal phase is stable, the low energy scaling of the density of states remains $\rho(E)\sim v^{-3}E^2$  and we can extract the effective velocity from $\rho''(0)\propto 1/v^3$. At finite energy, the quasiperiodic potential opens gaps, which forms minibands that contribute to the renormalization of the velocity.
 On approach to the transition as shown in Ref.~\cite{Pixley-2018}, $\rho''(0)$ diverges signalling a non-analytic density of states and the velocity goes to zero like $v \sim |W-W_c|^{\beta/3}$ with $\beta \approx 2$. In the diffusive metal phase, level statistics are random matrix theory like and plane-wave eigenstates at the Weyl node delocalize in momentum space. Extending these models to two dimensions shows that these transitions survive in the form of a semimetal to metal phase transition. By applying perturbation theory from twisted bilayer graphene yields a magic-angle condition in the velocity that coincides with the eigenstate phase transition. Thus,
rather remarkably, this transition is fundamentally related to the magic-angle effect in twisted bilayer graphene.

\section{Acknowledgements} 
We thank  Yang-Zhi Chou,  Sankar Das Sarma, Sarang Gopalakrishnan, Pallab Goswami, David Huse, Junhyun Lee, Rahul Nandkishore,  Leo Radzihovsky, Gil Refeal,  and Jay Sau for  various collaborations and discussions related to the work reviewed here, and we thank Bitan Roy, Bj\"orn Sbierski, and Sergey Syzranov for useful comments on an early draft as well as insightful discussions. We would also like to thank  Alexander Altland, Peter Armitage, Alexander Balatsky, Michael Buchhold, Sudip Chakravarty, Matthew Foster, Victor Gurarie, Silke Paschen, Thomas Searles, and Weida Wu for useful discussions.
This work is supported by NSF CAREER Grant
No. DMR-1941569.
The authors acknowledge the Beowulf cluster at the Department of Physics and Astronomy of Rutgers University, and the Office of Advanced Research Computing (OARC) at Rutgers, The State University of New Jersey (http://oarc.rutgers.edu), for providing access to the Amarel cluster and associated research computing resources that have contributed to the results reported here.
The Flatiron Institute is a division of the Simons Foundation.

\bibliographystyle{elsarticle-num}
\bibliography{RareRegionWeylReview}

\end{document}